\documentclass[journal]{IEEEtran}

\usepackage[latin1]{inputenc}
\usepackage{times,amsmath}
\usepackage{amssymb}
\usepackage{steinmetz}
\usepackage{pstool}
\usepackage{multirow}
\usepackage{enumerate}
\usepackage{graphicx}
\usepackage{booktabs}
\usepackage{subcaption}
\usepackage{mwe}
\usepackage{bigints}
\usepackage{MnSymbol}
\usepackage{stfloats} 
\usepackage[table]{xcolor}
\usepackage[square, comma, sort&compress, numbers]{natbib}
\usepackage{nohyperref}
\usepackage{algorithm,algorithmic}
\usepackage{epstopdf}
\usepackage{mathtools, cuted}

\usepackage{amsfonts}
\usepackage{multicol}
\usepackage{diagbox}
\usepackage{booktabs}
\usepackage{array}

\graphicspath{ {Figures/} }

\begin{document}

\title{ Non-contact Lung Disease Classification via OFDM-based Passive 6G ISAC Sensing }
%\title{ OFDM-based Passive ISAC Sensing in Microwave Band for Lung Disease Classification }
%\title{Non-Contact Lung Disease classification using RF Data Collected off the Chest}

\author{
\IEEEauthorblockN{
Hasan Mujtaba Buttar\IEEEauthorrefmark{1}, Muhammad Mahboob Ur Rahman\IEEEauthorrefmark{1}, Muhammad Wasim Nawaz\IEEEauthorrefmark{2}, Adnan Noor Mian\IEEEauthorrefmark{3}, Adnan Zahid\IEEEauthorrefmark{4}, Qammer H. Abbasi\IEEEauthorrefmark{5},\IEEEauthorrefmark{6} }

\IEEEauthorblockA{\IEEEauthorrefmark{1}Electrical engineering department, Information Technology University, Lahore 54000, Pakistan.\\ 
\IEEEauthorblockA{\IEEEauthorrefmark{3}Computer science department, Information Technology University, Lahore 54000, Pakistan.\\ 
\IEEEauthorrefmark{2}Computer engineering department, University of Lahore, Lahore, Pakistan.\\
\IEEEauthorrefmark{4}School of Engineering \& Physical Sciences, Heriot-Watt University, UK.\\
\IEEEauthorrefmark{5}James Watt School of Engineering, University of Glasgow, Glasgow, G12 8QQ, UK.\\
\IEEEauthorrefmark{6}Artificial Intelligence Research Center (AIRC), Ajman University, Ajman, UAE.\\
\IEEEauthorrefmark{1}\{mahboob.rahman\}@itu.edu.pk, \IEEEauthorrefmark{5}Qammer.Abbasi@glasgow.ac.uk }
}
\thanks{This work is supported in part by the UK EPSRC grants, CHEDDAR EP/X040518/1 and CHEDDAR Uplift EP/Y037421/1.}
}

% a.zahid@hw.ac.uk

\maketitle

\begin{abstract}

This paper is the first to present a novel, non-contact method that utilizes orthogonal frequency division multiplexing (OFDM) signals (of frequency 5.23 GHz, emitted by a software defined radio) to radio-expose the pulmonary patients in order to differentiate between five prevalent respiratory diseases, i.e., Asthma, Chronic obstructive pulmonary disease (COPD), Interstitial lung disease (ILD), Pneumonia (PN), and Tuberculosis (TB). The fact that each pulmonary disease leads to a distinct breathing pattern, and thus modulates the OFDM signal in a different way, motivates us to acquire OFDM-Breathe dataset, first of its kind. It consists of 13,920 seconds of raw RF data (at 64 distinct OFDM frequencies) that we have acquired from a total of 116 subjects in a hospital setting (25 healthy control subjects, and 91 pulmonary patients). Among the 91 patients, 25 have Asthma, 25 have COPD, 25 have TB, 5 have ILD, and 11 have PN. We implement a number of machine and deep learning models in order to do lung disease classification using OFDM-Breathe dataset. The vanilla convolutional neural network outperforms all the models with an accuracy of 97\%, and stands out in terms of precision, recall, and F1-score. The ablation study reveals that it is sufficient to radio-observe the human chest on seven different microwave frequencies only, in order to make a reliable diagnosis (with 96\% accuracy) of the underlying lung disease. This corresponds to a sensing overhead that is merely 10.93\% of the allocated bandwidth. This points to the feasibility of 6G integrated sensing and communication (ISAC) systems of future where 89.07\% of bandwidth still remains available for information exchange amidst on-demand health sensing. Through 6G ISAC, this work provides a tool for mass screening for respiratory diseases (e.g., COVID-19) at public places. 

%Finally, this work may also help to elevate the understanding of the clinicians about the pulmonary disease mechanism/pathology. 

%The results obtained are promising, and indicate the viability of the proposed method for seamless monitoring and identification of covid19 and other infectious diseases as well. 

\end{abstract}


\section{Introduction}
\label{sec:intro}

Respiratory diseases are caused by a number of risk factors, e.g., environmental factors (i.e., air pollution), non-healthy life-style (e.g., tobacco smoking), long-term exposure to allergens (e.g., irritating particles or gases, workplace fumes), bacterial infections, physical exertion, stress, and genetic factors. Respiratory diseases are a critical global health concern, and put a significant burden on healthcare sector in lieu of health insurance costs worldwide. The world health organization (WHO), in its 2019 report, identified respiratory diseases as the third-largest cause of mortality globally, behind cardiovascular diseases and cancer, with approximately 3.8 million annual deaths attributed to them \cite{world2020top}. 

Respiratory diseases fall into two distinct categories: obstructive and restrictive lung diseases, each significantly impacting lung function and breathing capacity in unique ways. Obstructive lung diseases involve a narrowing or obstruction of the airways, impeding the proper flow of air into and out of the lungs. This restriction primarily affects exhalation, resulting in airflow limitations \cite{cho2020aging}. In contrast, restrictive lung diseases involve stiffness or damage to the lung tissue itself, impairing lung expansion and reducing lung volume.

This work aims to diagnose the following five pulmonary diseases in a contactlass manner: Asthma, Chronic obstructive pulmonary disease (COPD), Interstitial lung disease (ILD), Pneumonia (PN), and Tuberculosis (TB). Further, since this work utilizes radio sensing modality that captures the changes in the breathing pattern to differentiate between various diseases, we consider it imperative to first do a brief discussion about the pathophysiology of the each of these five lung diseases from the perspective of how it impacts the breathing pattern.

\subsection{Brief pathophysiology of the lung diseases} 
We begin with the observation that a healthy individual typically exhibits a normal respiratory rate, full, deep breaths, and a relaxed chest wall during breathing.
Next, below we outline the pathophysiology of each of five pulmonary diseases, followed by its prevalence, followed by our remarks on how it impacts the breathing pattern.
\begin{itemize}
\item {\it Asthma:} Asthma is a obstructive lung disease, is characterized by airway inflammation and constriction. Asthma affects a significant global population, with an estimated 235 million individuals currently afflicted worldwide, particularly impacting children, among whom it affects 14\% and continues to rise \cite{padilla2018lung}. The breathing pattern of a person with asthma often includes wheezing, dyspnea, increased respiratory rate, and use of accessory muscles.
\item {\it Chronic obstructive pulmonary disease (COPD):} COPD belongs to the obstructive category, representing a progressive and chronic condition comprising chronic bronchitis and emphysema \cite{marccoa2018classification}. The prevalence of this condition is increasing, with a rate of $11.7\%$, resulting in roughly 3 million fatalities per year \cite{world2020top}. The breathing pattern of a person with COPD often includes dyspnea, tachypnea, pursed-lip breathing, use of accessory muscles, and barrel chest.
\item {\it Interstitial Lung Disease (ILD):} ILD causes inflammation and scarring of the lung's interstitium, hindering oxygen transfer into the bloodstream. In 2019, approximately 2.28 million men and 2.43 million women globally suffered from ILD \cite{zeng2023global}. The breathing pattern of a person with ILD often includes tachypnea, dyspnea, shallow breathing, and use of accessory muscles.
\item {\it Pneumonia (PN):} PN is an infection of lung tissue, can also contribute to reduced lung expansion due to inflammation and fluid buildup. Notably, PN has long ranked among the top three causes of death and disability worldwide, affecting both children and adults \cite{marciniuk2014respiratory}. The breathing pattern of a person with PN often includes tachypnea, shallow breathing, increased work of breathing, and dyspnea.
\item {\it Tuberculosis (TB):} Tuberculosis afflicts approximately 10.4 million people annually, resulting in an estimated 1.4 million deaths in 2015 alone \cite{world2013global}. The breathing pattern of a person with tuberculosis often includes cough, dyspnea, rapid and shallow breathing, chest pain, wheezing or crackles, and hemoptysis.
\end{itemize}

%Chronic obstructive pulmonary disease (COPD), asthma, acute respiratory infections, tuberculosis (TB), lung cancer, and recent COVID-19 infection are among the most common respiratory diseases associated with severe illness and mortality worldwide. COPD is characterized by progressive airflow restriction due to chronic inflammation in the airways and lungs, contributing to its high morbidity and mortality rates, thus making it the third leading cause of death globally. Asthma is characterized by the narrowing and swelling of airways, often accompanied by the production of excess mucus. These symptoms can lead to difficulties in breathing, coughing, wheezing (a whistling sound when exhaling), and shortness of breath. Globally, an estimated 235 million individuals currently suffer from asthma, making it one of the most prevalent chronic diseases. Notably, asthma affects approximately 14\% of children worldwide, rendering it the most common chronic childhood ailment. Each year, over 4 million individuals succumb to respiratory infections, with acute lower respiratory tract infections such as pneumonia and acute bronchitis persistently ranking among the top three causes of mortality and disability across all age groups. Pneumonia, caused by viruses, bacteria, or fungi, represents a significant pulmonary abnormality contributing to this burden. 

Respiratory illnesses are treatable with early detection and proper prevention measures. However, variations in diagnostic standards contribute to the high mortality rates associated with pulmonary diseases. The diagnosis heavily relies on the expertise of medical professionals, and thus, misdiagnosis could lead to irreversible harm to patients.

\subsection{The gold standard for lung disease diagnosis}
The gold standard for clinical diagnosis of respiratory diseases is a complex, multi-pronged process, which consists of the following sensing modalities: 
\begin{itemize}
\item {\it Clinical assessment:} A thorough patient history and physical examination remain fundamental. Symptoms, exposure history, and physical signs guide the choice of further diagnostic testing.
\item {\it Stethoscope-based auscultation analysis:} 
Medical professionals use stethoscopes to listen to lung sounds, which are generated by the passage of air through the lungs during respiration. These sounds are crucial for detecting respiratory diseases. For example, Asthma and COPD can cause airway obstruction, leading to abnormal lung sounds such as wheezes and crackles. 
\item {\it Medical imaging techniques:} Chest X-rays and CT-scans are critical for diagnosing many respiratory diseases, including pneumonia, tuberculosis, and lung cancer \cite{al2023multi}. High-resolution imaging provides detailed information on the lung's condition, helping to identify abnormalities.
\item {\it Pulmonary function tests (PFTs):} These tests measure the lungs' capacity to hold and expel air and how efficiently they transfer oxygen into the blood. Spirometry is the most common PFT and is essential for diagnosing conditions like Asthma and COPD. Spirometry measures the amount of air one can forcefully exhale and the speed at which one exhales it \cite{gordon2012spirometry}.
\item {\it Laboratory tests:} Blood tests, biomarkers, and sputum culture analysis can identify infections and inflammatory conditions. For example, sputum culture involves collecting and analyzing a sample of phlegm (mucus coughed up from the lungs) to identify bacteria or other pathogens that might be causing a respiratory infection. The predominant categories of pathogenic bacteria identified in a sputum culture encompass those responsible for the development of Pneumonia and Tuberculosis \cite{anevlavis2009prospective,nguyen2019factors}. Similarly, \cite{mcdowell2022machine} investigates the relationship between circulating microbial extracellular vesicles (EVs) extracted from patient serum (blood) and respiratory diseases, with the goal of classifying COPD, asthma, and lung cancer using machine learning techniques.

\item {\it Endoscopic procedures:} Bronchoscopy allows direct visualization of the airways and biopsy for histopathological examination, which is crucial for diagnosing pulmonary diseases.
\end{itemize}

However, we note that the traditional diagnostic methods are costly, time-consuming, invasive, and unsuitable for daily screening (e.g., X-ray and CT scans). Therefore, there is an urgent need to develop novel non-invasive, non-contact methods that could enable daily screening, rapid diagnostics, and early detection of pulmonary diseases. Such methods are particularly appealing because they could reduce the patient discomfort and risk of infection.

%5. Artificial Intelligence and Machine Learning: AI and ML are being integrated with existing diagnostic methods (like imaging) to enhance accuracy and speed in diagnosing respiratory diseases, potentially even identifying patterns not evident to human observers.

%These innovations represent the forefront of research and development in the field. However, the integration of non-invasive and non-contact methods into clinical practice varies widely depending on the technology's maturity, cost, availability, and the specific healthcare context. Continuous validation and standardization of these methods are necessary to ensure their reliability and accuracy for clinical use.

\subsection{Contributions} 
Inline with recent research trends, this paper presents a novel non-contact method that utilizes orthogonal frequency division multiplexing (OFDM) signals in the microwave band to radio-expose the pulmonary patients in order to differentiate between five kinds of respiratory diseases, i.e., Asthma, COPD, ILD, PN, and TB. 
Specifically, the key contributions of this work are as follows:
\begin{itemize}
    \item We acquire OFDM-Breathe dataset, first of its kind. It consists of 13,920 seconds of raw RF data (at 64 distinct OFDM frequencies in the microwave band) that we have acquired from a total of 116 subjects in a hospital setting (25 healthy control subjects, and 91 pulmonary patients). Among the 91 patients, 25 have Asthma, 25 have COPD, 25 have TB, 5 have ILD, and 11 have PN.
    \item We implement a number of machine learning (ML) and deep learning (DL) models in order to do lung disease classification using OFDM-Breathe dataset. Among DL models, an overall test accuracy of 98\% is achieved by the convolutional neural network (CNN) model, while an accuracy of 97\% is achieved by the long short-term memory (LSTM) and Transformer models. The CNN model turns out to be the winner in terms of precision, recall, and F1-score. Among the ML models, the multi-layer perceptron (MLP) model along with its variants stands out with an overall accuracy that is $>95.6$\%.
    \item The ablation study reveals that it is sufficient to radio-observe the human chest on seven different microwave frequencies out of a total of 64 distinct OFDM frequencies, in order to make a reliable diagnosis (with 96\% accuracy) of the underlying lung disease. This corresponds to a sensing overhead that is merely 10.93\% (7/64*100) of the allocated bandwidth. This points to the feasibility of 6G integrated sensing and communication (ISAC) where about 89.07\% of bandwidth still remains available for information exchange amidst on-demand health sensing. 
\end{itemize}

{\it To the best of our knowledge, this is the first work that utilizes a handful of microwave frequencies for lung disease classification in a reliable and contactless manner.}
Additionally, this work is also a contribution to the themes of 6G ISAC, smart homes, smart cities, and digital twin. 

{\bf Rationale.}
We utilize an OFDM signal consisting of a set of microwave frequencies (sub-carriers) that gets modulated by the rhythmic (or arrhythmic) movements of the chest of the pulmonary patients due to pathological respiratory performance of their lungs. Each lung disease affects the inhalation-exhalation pattern of the lungs in a different way, that is captured by the OFDM sub-carriers. Further, each OFDM sub-carrier records distinctive information about the underlying lung disease due to unique frequency, phase and amplitude modulation it goes through. The modulated OFDM signal is translated to channel frequency response (CFR) which when passed to AI methods, allow us to identify the underlying lung disease in a contactless manner.

\subsection{Outline} 
The rest of this paper is organized as follows. Section II discusses the selected related work. Section III provides a compact discussion of the experimental setup that we have utilized to acquire the modulated CFR data from the chests of pulmonary patients in a contactless manner. Section III also provides the specifics of the data acquisition protocol we have utilized in order to construct our custom dataset. Section IV summarizes the key data pre-processing steps. Section V provides a brief discussion of the various ML and DL classifiers we have implemented. Section VI does detailed performance analysis and section VII concludes the paper.

%this work doesn't discuss covid19 and lung cancer, due to difficulty in acquiring such data.

% {\bf Notations.} Unless specified otherwise, $\vert.\vert$ and $\Vert.\Vert$  denote the modulus and the 2-norm respectively, $\mathbb{E}(.)$ is the expectation operator, boldface letters such as $\mathbf{X}$ represents a vector and $\mathcal{CN}$ means complex normal.
% \begin{figure}[htp]
%     \centering
%     \includegraphics[width=6cm]{Updated_system_final1.jpg}
%     \caption{System Model}
%     \label{fig:sys}
% \end{figure}

\section{Related work}

This section summarizes selected related work on {\it non-invasive and non-contact} lung disease classification that ranges from medical imaging (X-ray, CT scan)-based methods, to acoustics (cough, breathing sound, lungs sound)-based methods, to radio signals (radar, SDR)-based methods. We note that a vast majority of these works rely upon a range of artificial intelligence (AI) techniques in order to do rapid COVID-19 and lung cancer detection in particular, and lung disease classification in general (see the survey articles \cite{shi2020review,binczyk2021radiomics,qin2018computer} and references therein). We also outline some miscellaneous methods, e.g., chemical analysis of exhaled breath \cite{kim2021recent}, etc.

%chest X-ray and computed tomography (CT) scans based methods:

\subsection{Medical imaging-based methods} 
The diagnosis of lung diseases using medical imaging data, e.g., chest CT scan, chest X-ray images, etc., has traditionally required the expertise of a radiologist or a physician. However, the landscape has drastically changed recently whereby AI techniques have shown lots of promise for automatic respiratory disease classification using medical imaging data \cite{kieu2020survey}. In fact, automatic COVID-19 disease diagnosis by doing AI-based analytics on various medical imaging modalities has been one of the most researched problem in the post-COVID-19 era \cite{mei2020artificial}. Among such works, a fair amount of works investigate the feasibility of COVID-19 detection and diagnosis of other lung diseases using chest X-rays images only, owing to the fact that it is a less expensive, faster, and widely available technology \cite{koyyada2024systematic}, \cite{wang2021deep}. Moving forward, we note that though the related work on lung disease diagnosis using medical imaging data is abundant, nevertheless, we discuss only a few related works below, due to the space constraints. 

%Diseases like Pneumonia, COVID-19, and Tuberculosis can be detected by analyzing patient’s chest X-ray within a short time with trustable accuracy using a deep learning technique \cite{mekov2020artificial}.

Authors of \cite{chandra2020pneumonia} acquire 412 chest X-ray images of Pneumonia patients, segments them to obtain various lung regions in order to extract eight statistical features, and achieves a remarkable accuracy of 95.39\% using the MLP classifier. 
\cite{zhang2021pneumonia} utilizes chest CT scan images of 52 Pneumonia patients, extracts six features, utilizes a CNN, and reports an accuracy of 97\%.
\cite{malik2023cdcnet} considers a multi-clas classification problem whereby the aim is to differentiate between COVID-19, Pneumothorax, Pneumonia, lung cancer, and Tuberculosis. To this end, they apply fuzzy tree transformation to the X-ray images, do feature extraction, and implement three CNN-based pre-trained models, i.e., VGG-19, ResNet-50, and Inception v3 which achieve an accuracy of 95.61\%, 96.15\%, and 95.16\%, respectively. 
\cite{tuncer2021novel} considers a small dataset consisting of chest X-ray images of COVID-19 and Pneumonia patients, applies the fuzzy tree transformation to the images, do feature extraction using a multikernel local binary pattern, and implement traditional machine learning classifiers which achieved an accuracy of 97.01\%.
\cite{deng2024pneumonia} considers a dataset comprising 3,345 chest X-ray images of Pneumonia patients, implements a CNN that achieves an accuracy of 97.64\%, and develops a PneumoniaApp with the hope that it helps physicians localizing the regions of lung opacity. \cite{al2023multi} enhances the quality of CT scan and X-ray images using k-symbol Lerch transcendent functions. Subsequently, two pre-trained deep learning models, AlexNet and VGG16Net, were developed to distinguish between three different lung conditions: pneumonia, COVID-19, and a healthy lung. Following image enhancement, the models achieved outstanding accuracies of 98.60\% and 98.50\% for the X-Ray and CT scan datasets, respectively.

While the researchers have largely remained focused on early computer-aided detection of COVID-19, lung cancer and Pneumonia using medical imaging data, we note that the research on AI-empowered diagnosis of Asthma, COPD \cite{kaplan2021artificial}, ILD \cite{mei2023interstitial}, and TB \cite{harris2019systematic} using CT scan and X-ray images is also picking up lately. 

% we don't discuss works on lung cancer detection, as it is out of scope of our paper.
%\cite{gao2024robust} design FULFIL Forecasting Uncompleted Labels For Inexpensive Lung nodule detection algorithm that plays a significant role in the detection of pulmonary nodules in low-dose computed tomography (LDCT) scans, contributing to the diagnosis and treatment of lung cancer.

%This breakthrough not only showcases the potential of AI in healthcare but also highlights the significance of collaborative efforts between technology and medical professionals in building innovative solutions for better patient care.

\subsection{Respiratory sounds-based methods}
Lung sounds exhibit significant changes due to different respiratory disorders. For example, Asthma causes wheezing that results from heavy mucus production or airway inflammation. Wheezing sounds is typically heard during expiration but sometimes during inspiration. Wheezing and other such abnormal sounds could be heard when a sick person breathes or speaks. 
This has motivated the researchers to design a collection of methods that utilize one of the following modalities to diagnose COVID-19 and other lung diseases: the lungs sounds (through auscultation), speech signals, breathing sounds, and cough sounds \cite{aly2022pay,brown2020exploring}. Below, we discuss the three acoustic methods, one by one. 

\begin{itemize}
\item {\it Lung auscultation-based methods:}
This method entails measuring lung sounds using an electronic stethoscope. \cite{choi2023interpretation} does lung auscultation to record wheezing and crackle sounds of 126 pulmonary patients in a hospital setting, and implement an improved VGG network by incorporating a lightweight attention mechanism to diagnose various lung diseases (i.e., Asthma, COPD, ILD, Bronchiectasis, and Pneumonia).  
\cite{zhang2024research} utilizes the ICBHI2017 challenge lungs dataset, extracts MFCC features, and implements a dual-channel CNN-LSTM for diagnosis of various respiratory diseases (i.e., COPD, Pneumonia, Bronchiectasis, URTI, and Bronchiolitis).
\item {\it Voice-based methods:}
This method involves capturing high-fidelity speech recordings via a microphone. \cite{boyanov1997acoustic} argues that laryngeal diseases could be diagnosed based on the noticeable change in vocal tracts that leads to a pathological voice. \cite{nallanthighal2019deep} utilizes CNN and RNN model for breathing event detection while during conversation, with the aim to detect COPD, asthma, and general respiratory infections. 
\item {\it Cough-based methods:} 
This method entails recording high-fidelity cough sounds using smartphones or microphones. Cough-based diagnosis of various respiratory diseases such as Asthma, Tuberculosis, Bronchitis, etc., through signal processing and AI-based analytics has attracted considerable attention recently \cite{ijaz2022towards}.
\cite{abeyratne2013cough} acquires Pneumonia coughs, Asthma coughs, and Bronchitis coughs of children through microphones, and performs time-series statistical analysis, and time-frequency analysis in order to reliably detect childhood Pneumonia. 
\cite{swarnkar2013automatic} applies spectral analysis, time-series analysis and statistical methods with the aim to differentiate between the wet and dry coughs of children. 
\cite{al2013signal} does frequency-domain analysis of the cough of Asthmatic patients and concludes that it is distinct from the cough of non-Asthmatic patients as it has higher energy in low-frequency components. 
Finally, \cite{imran2020ai4covid},\cite{xiong2022reliability} implements a CNN and RNN model in order to do COVID-19 detection based on cough.
\end{itemize}

\subsection{RF-based methods}
RF-based methods consist of radar and software-defined radio (SDR)-based methods, and are discussed below, one by one.

\begin{itemize}
\item {\it Radar-based methods:} 
Radar-based methods utilize the classical range-Doppler principles in order to localize and track the objects of interest, i.e., rhythmic movement of human chest in case of (pulmonary and cardiac) health sensing \cite{park2019preclinical}. 
\cite{li2022detection} does joint estimation of the position of a person and his/her breathing rate in a non-line-of-sight scenario using a 77-GHz frequency-modulated continuous-wave (FMCW) radar. 
\cite{van2016wireless} performs a diagnostic cross-sectional study on eight postoperative patients whereby a radar system is used to measure their respiratory rate during mechanical ventilation during spontaneous breathing. 
Authors of \cite{yu2019highly} design a novel ultrasonic-based radar with desirable characteristics (e.g., highly linear, phase-canceling, self-injection-locked) which could reliably and simultaneously measure both the breathing rate and the heart rate of a person. \cite{beltrao2022contactless} elegantly monitors the delicate breathing signals of premature infants within the neonatal intensive care unit (NICU) using a sophisticated radar system operating in the 24-GHz ISM (industrial, scientific, and medical) band.
Finally, authors of \cite{ha2020contactless} aim to monitor the performance cardiac system whereby they radio-expose the human chest with radar signals, and utilize generative AI methods to construct a seismocardiograph signal.
\item {\it SDR-based methods:}
The SDR-based health sensing methods monitor the human respiratory system based on small chest movements that are recorded through the modulated OFDM signals reflected off the human chest. 
In \cite{ashleibta2020non}, \cite{khan2021intelligent}, authors utilize the CFR data acquired-off the human chest in order to detect and classify three breathing patterns: normal, shallow and elevated breathing.
\cite{rehman2021improving} extends this further by classifying up to eight breathing patterns, i.e., eupnea, bradypnea, tachypnea, Biot, sighing, Kussmaul, Cheyne-Stokes, and central sleep apnea. 
Kawish et. al. \cite{kawish2023hand} adopt a different approach where instead of the chest, they radio-expose the hand in order to differentiate between three kinds of breathing patterns, i.e., Tachypnea, Bradypnea and Eupnea. 
\cite{santos2024non} aims to synthesize a PPG signal using generative AI methods by utilizing the modulated OFDM signal reflected-off the chest of the chest of a healthy subject.
Finally, \cite{buttar2023non} radio-observes the chest of a subject in order to infer his/her hydration status in a non-contact manner. 
\end{itemize} 

\subsection{Other Methods}

There exist some other non-contact sensing modalities as well that could aid in diagnosis of pulmonary diseases. For example, exhaled breath analysis methods assess the chemical composition of exhaled breath using sensors and spectrometry, and can detect biomarkers for diseases like asthma, COPD, and lung cancer \cite{kim2021recent}. Another method is imaging with infrared thermography which could assess respiratory function by detecting temperature variations related to breathing patterns and nasal flow. Pulse oximetry is yet another non-invasive method that measures Oxygen saturation in the blood and can be a quick indicator of respiratory function, though not specific for diagnosis \cite{AhsanECGpaper}.

\subsection{Research gap}

{\it This work is closest in scope to the radar and SDR based methods which radio-expose the human chest with wideband RF signals in order to perform various health sensing tasks. To the best of authors' knowledge, there is no documented instance in the accessible literature of employing Software Defined Radio (SDR) for non-contact classification of lung diseases. Instead, existing approaches typically focus on classifying breathing patterns by observing various respiratory actions in healthy individuals, rather than utilizing microwave (OFDM) signals to classify lung diseases in diagnosed patients, to date, which is precisely the agenda of this work.} 
Additionally, this work also contributes to the themes of passive but event based 6G ISAC for health sensing, smart homes, smart cities, and digital twin.

\section{Experimental Setup \& Data Acquisition}
\label{sec:method}

%This section provides sufficient details about the hardware and software setup used to construct the custom dataset, our thoughtful data collection methodology (that helped us capture data in a very controlled manner), as well as the intricate details of the two experiments performed in order to collect data for the proposed method. 

\subsection{The experimental setup}
\label{sec:sys-model}

Our experimental setup for non-contact monitoring and identification of pulmonary diseases consists of two SDRs, basically, Universal Software Radio Peripheral (USRP) model B210 SDRs. We use MATLAB R2021a to program both the transmit and receive USRP SDRs. The two radios communicate by exchanging OFDM signals between them. Some pertinent details of the transmit USRP SDR and the receive USRP SDR are as follows. 

\begin{figure}[t]
\begin{center}
	\includegraphics[width=9cm,height=6cm]{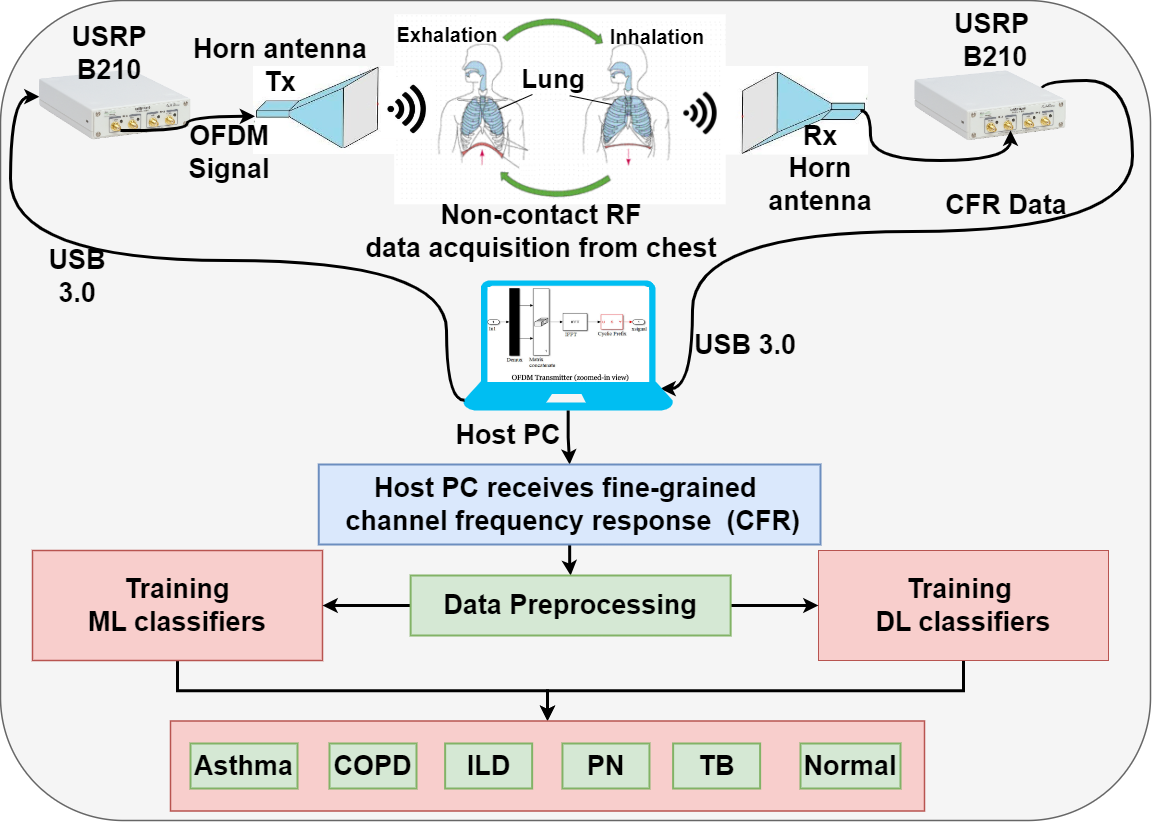} 
\caption{{The proposed method: An SDR-based OFDM transceiver illuminates the chest of a subject with radio waves, collects the reflected signals, passes them to various machine and deep learning methods, which ultimately classify a subject either as healthy or sick with a pulmonary disease.}}
\label{fig:sysmodel}
\end{center}
\end{figure}

{\it USRP SDR-based OFDM Transmitter:} 
We group the data in batches of 128 bits, and map them to quadrature phase-shift keying (QPSK) symbols. We then take the 64-point inverse fast Fourier transform (IFFT) of the vector containing QPSK symbols, and append a cyclic prefix (CP) of 16 samples to set the length of the OFDM frame to 80 samples. Finally, a 40 dBi antenna is utilized to transmit the OFDM signal.
% The gain of the transmit antenna is set to 40 dBi.

{\it USRP SDR-based OFDM Receiver:} 
Having received an OFDM frame, we first remove the CP, and then take the 64-point fast Fourier transform (FFT). Next, we compute the channel gain $h_i$ on $i$-th sub-carrier as follows: $h_i=\frac{y_i}{x_i}$, where $x_i$,$y_i$ is the sent and received QPSK symbol on $i$-th sub-carrier, respectively. Note that $x_i,y_i,h_i\in \mathbb{C}$. The channel frequency response $\mathbf{h}=[h_1,\cdots,h_N]^T$ constitutes the raw RF data, which we later feed to our ML and DL algorithms in order to classify each subject as either healthy or sick with a pulmonary disease.

Fig. 1 provides a pictorial overview of the methodology of this paper and TABLE \ref{table:usrp_param} summarizes the values of the key parameters that we have used to configure the transmit and receive USRP SDRs.
% Additionally, Fig. 1 provides a pictorial overview of the methodology of this paper. 

% \begin{figure*}
% \hfill
% \subfigure[The transmitter flowgraph]{\includegraphics[scale=0.42]{Figures/MAtlab_simulink.Tx.drawio (5).png}}
% \hfill
% \subfigure[The receiver flowgraph]{\includegraphics[scale=0.42]{Figures/MAtlab_simulink.Tx.drawio (5).png}}
% \hfill
% \caption{The Simulink flowcharts of the USRP-SDR based OFDM transmitter and receiver. }
% \label{fig:flowg}
% \end{figure*}

%needs to populate it.
\begin{table}[t]
\centering
\begin{tabular}{|c| c|} 
 \hline
 
 {Parameter} & {Type/Value}
	
\\
\hline

Modulation scheme &	QPSK \\
No. of OFDM subcarriers &	$64$\\
Size of FFT/IFFT &	$64$ points\\
Size of cyclic prefix	& $16$\\
Sampling rate&	$1000$ samples/sec\\
Centre frequency&	$5.23$ GHz\\
Gain (at Tx \& Rx) &	$40$ dB\\

 \hline
\end{tabular}
\caption{{Important parameters for the OFDM transceiver used for non-contact identification of lung diseases.}}
\label{table:usrp_param}
\end{table}

\subsection{The OFDM-Breathe dataset}

Under the proposed method, the transmit SDR exposes the chest of the subject to the OFDM signal, while the receive SDR collects the signal reflected off the chest of the subject\footnote{This study was approved by the ethical institutional review board (EIRB) of Information Technology University (ITU), Lahore, Pakistan, as well as from the EIRB of the local hospital in Lahore. Each subject (healthy or patient) read and signed the informed consent form. Finally, as per approved protocol, we only collected data from those pulmonary patients who were in clinically stable condition. }. 
This lets us acquire our custom dataset---the so-called {\it OFDM-Breathe dataset}, which we believe is first of its kind. 

{\it The significance of the OFDM-Breathe dataset:}
To the best of our knowledge: i) OFDM-Breathe dataset is the {\it first} labeled dataset that captures {\it five kinds of respiratory disorders (i.e., Asthma, COPD, ILD, PN, TB) in actual patients} via OFDM signals in the microwave band; ii) OFDM-Breathe dataset is the {\it largest} among all previously reported datasets that utilize OFDM signals in the microwave band for various health sensing tasks. Specifically, OFDM-breathe dataset consists of 13,920 seconds of raw RF data (at 64 distinct OFDM frequencies in the microwave band) that we have acquired from a total of 116 subjects (both healthy control group, and patients diagnosed with five kinds of respiratory diseases). iii) {OFDM-Breathe dataset is collected in the city of Lahore, Pakistan during the smog season (i.e., from October 2023 to January 2024). The city of Lahore consistently remained among one of the most polluted cities in the world throughout the year 2023 in terms of PM2.5, PM10 concentration, and air quality index \cite{majeed2024solving}}. Thus, the {\it OFDM-Breathe dataset could provide valuable insights into the aggregate impact of extreme environment (pollution) on respiratory health of the civic population of Lahore}.

{\it Data description:}
Following comments about the OFDM-Breathe dataset are in order. i) This is a cross-sectional study whereby the dataset consists of 84 male subjects with an average age of 51.5 years, and 32 female subjects with an average age of 37.3 years. ii) The dataset comprises 25 subjects from the healthy control group, with the remaining 91 subjects being formally diagnosed patients experiencing an active onset of one of the following respiratory diseases: 1) Asthma, 2) COPD, 3) TB, 4) ILD, and 5) PN. iii) The data of the healthy control group is collected from healthy subjects (students and staff) at university main campus. iv) The data of the lung disease patients is collected from a local hospital in Lahore. v) For each of the five lung diseases, the labels are assigned by the on-duty clinician (who diagnosed the disease using clinical symptoms, and chest X-rays). Further, data is collected from those admitted patients only who were stable and conscious. 
vi) Among the 91 patients diagnosed with a respiratory disease, 25 have Asthma, 25 have COPD, 25 have TB, 5 have ILD, and 11 have PN. 
Note that the OFDM-Breathe dataset is somewhat imbalanced, from the perspective of ILD and PN patients. 
%We gathered data four times for 30 seconds for each person. 

\begin{figure}[ht]
\begin{center}
	\includegraphics[width=7.5cm,height=4.5cm]{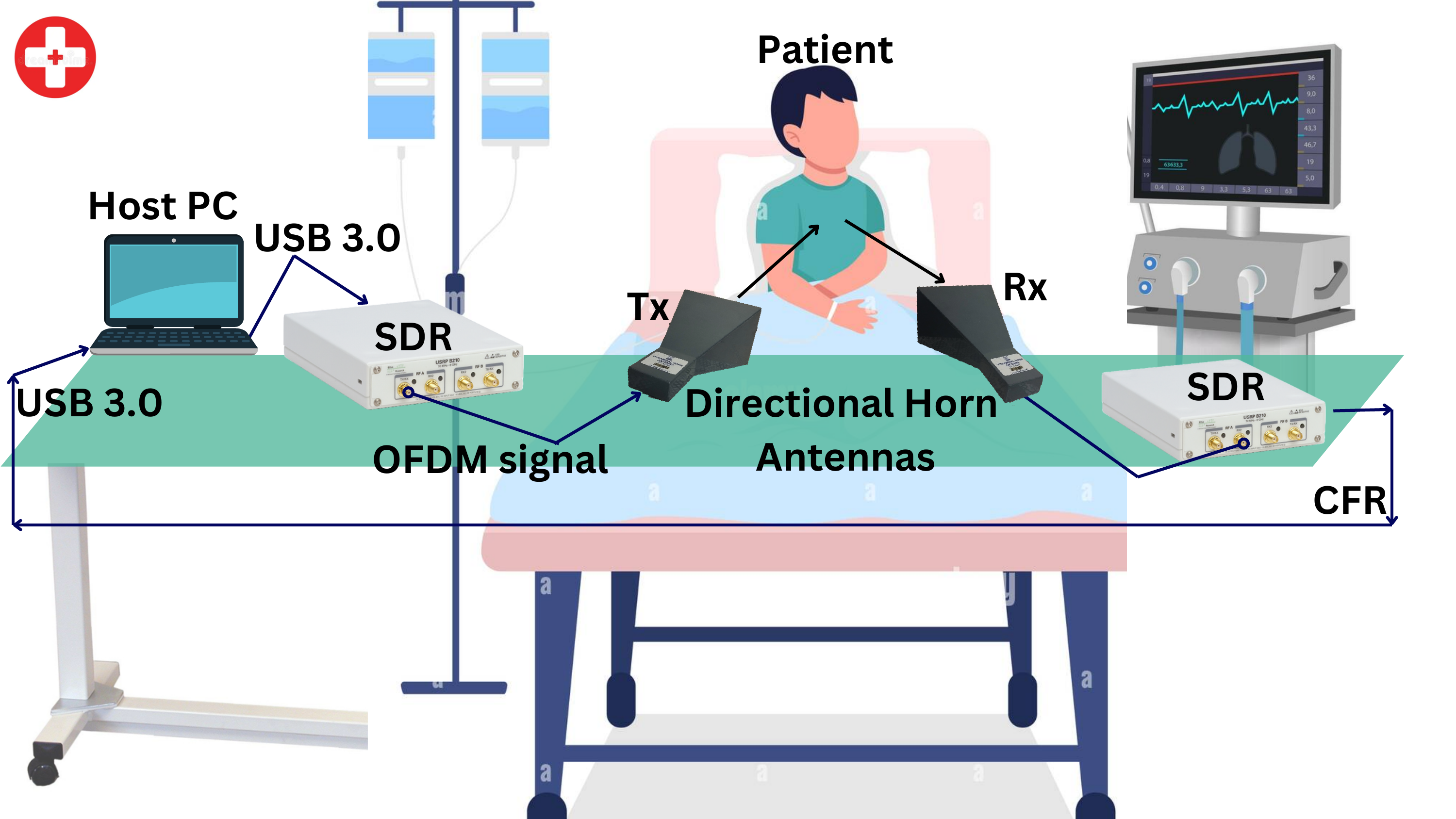} 
\caption{The experimental setup used for acquisition of the OFDM-Breathe dataset. It consists of two USRP B210 SDRs, two directional horn antennas, and a PC.}
\label{fig:exp}
\end{center}
\end{figure}

{\it Data collection protocol \& data statistics:}
Under proposed method for data collection, each participant sits on a chair that is roughly 2.5 feet away from a table that hosts the two USRP SDRs and the two directional horn antennas which both point towards the chest of the subject (see Fig. \ref{fig:exp}). The transmit antenna strikes an OFDM signal onto the chest of the subject, while the receive antenna collects the signals reflected off the subject's chest. For the duration of experiment (which is 30 seconds), the subject remains stationary in order to make sure that there are no motion artifacts in the collected data. For each subject, we repeat the experiment four times. This allows us to collect 50 minutes of data for each of the Asthma, COPD, TB, and normal classes, 10 minutes of data for the ILD class, and 22 minutes of data for the PN class. Overall, the OFDM-Breathe dataset consists of a total of 232 minutes of raw CFR data, due to a total of 464 (=116x4) experiments. 

%Dataset dimensions: 140600 rows × 14 columns (13 frequencies +1 disease)

%\textcolor{red}{We may want to add a table that quickly describes the key stats of our dataset.}

\section{Data pre-processing}

We do the following pre-processing steps in order to prepare the recorded CFR data for our ML/DL models:

\begin{itemize}
\item {\it Decimation-in-time:}
We first make the observation that the rhythmic inhalation and exhalation activity of the lungs (i.e., 12-20 breaths/min under sinus conditions, and {20-30 breaths/min under pathological conditions}) implies that the frequency band of interest lies in a narrow range of 0-1 Hz. This allows us to do decimation-in-time (DiT) by a factor of 10 in order to reduce the sampling rate of the CFR data to 100 Hz. 
\item {\it Decimation-in-frequency:}
Further, we have empirically observed that the human chest responds to the two adjacent OFDM sub-carriers $i,j$ in a similar way, i.e., $|h_i-h_j|<\delta_m$ and $\angle (h_i-h_j)<\delta_p$, $\forall i,j=1,...,N=64$ where $\delta_m,\delta_p$ are arbitrarily small numbers. This allows us to do decimation-in-frequency (DiF) by a factor of 4. Note that both DiT and DiF together help significantly reduce the memory and computational needs of the proposed method. 
\item {\it Data segmentation:}
Next, we do data segmentation whereby we split the CFR data corresponding to each recording session of duration 30 seconds into three smaller segments of duration 10 seconds each. Thus, size of each segment is: $\mathbf{H}\in \mathbb{C}^{16\times 1000}$. Further, we repeat the class label across all the three segments of that recording. This way, the size of the OFDM-Breathe dataset increases by three-fold. Specifically, we obtain 300 labeled examples for each of the Asthma, COPD, TB, and the normal classes, and 60 and 132 labeled examples for ILD and PN, respectively. 
\item {\it Denoising:}
We utilise the 8-level discrete Wavelet transform with db4 mother wavelet to denoise (low-pass filter) the segmented CFR data, and to remove the artifacts.  
\item {\it Data reshaping:} Finally, we reshape the segmented and denoised CFR data as per needs of various ML and DL classifiers. 
\end{itemize}

%\textcolor{red}{Further, recall that we gathered data from five subjects, and for each subject, we conducted five experiment sessions each of duration 30 seconds. Thus, for each class (hydrated/dehydrated), we accumulated raw CFR data which consists of 750,000 samples, i.e., 5 subjects * 5 experiments/subjects * 30 second/experiment  * 1000 samples/sec.  We partitioned this data into segments, with each segment consisting of 3000 samples. Thus, through this meticulous approach, we successfully compiled 250 examples for each of the "hydrated" and "dehydrated" classes.} 

%\begin{figure}[ht]
%\begin{center}
%	\includegraphics[width=9cm,height=3cm]{Figures/Synopsis Defence (1).png} 
%\caption{Experimental setup of the proposed HBDM method. }
%\label{fig:methodour}
%\end{center}
%\end{figure}

\section{Machine \& Deep Learning Classifiers Implemented }  

\subsection{Machine Learning Classifiers}
We use Matlab in order to feed the OFDM-Breathe dataset to the following ML classifiers and their variants: linear discriminant (LD), K-nearest neighbours (KNN), support vector machine (SVM), and ensemble classifier. Furthermore, we also train and evaluate three variants of a multi-layer perceptron (MLP), i.e., narrow, medium and wide MLP with a single hidden layer, but with 10, 25, 100 neurons, respectively. We also implement bi-layered MLP (tri-layered MLP) which has two (three) dense layers with 10 neurons in each layer. Finally, we use K-fold cross-validation strategy in order to avoid the over-fitting issues. 

%Subsequently, we provide detailed performance analysis and comparison of all the ML and DL classifiers implemented. 

\subsection{Deep Learning Classifiers}

We implement three DL models, namely, Transformer, LSTM and CNN, in Python, Keras and Tensorflow framework. 

\subsubsection{The Transformer model}

%Since there are 64 frequencies, we have subsampled them by a factor of 5 (20\%); thus, we have 13 frequencies as feature0, feature1, to feature12. three-way split. train-validation-test ratio of 75-15-15. 

The Vanilla Transformer model that we have implemented consists of an embedding layer, followed by a Transformer block, followed by pooling, dropout and dense layers. Together, all these layers help Transformer model learn the underlying representations of input RF sequences and perform the classification task at hand. Below, we provide a compact briefing about each layer of our Transformer model:
\begin{itemize}
\item {\it Embedding Layer:}
The input sequence goes through two separate embedding layers, one for tokens (which gives us a dense, fixed-length embedding vector for each token), and another for token indices, i.e., positions (which helps the model to learn the relative positions of tokens in the input sequence). Eventually, we combine token embeddings with positional embeddings. This layer is parameterized by the following: 1) {\it max-len} which is maximum length of input sequences, 2) {\it embed-dim} that specifies the size of the embedding vector for each token in the input sequence, 3) {\it vocab-size} which specifies the size of the vocabulary, i.e., the total number of unique tokens in the input data.
\item {\it Transformer Block:}
The Transformer block consists of three main components: 1) Query, Key, and Value vectors, 2) Attention scores, 3) Weighted sum of values. Specifically, each input token (or position) is first associated with three vectors: query, key, and value. Then, the self-attention mechanism computes attention scores between each query and key pair. Finally, the attention scores are used to compute a weighted sum of the corresponding value vectors.
This weighted sum represents the attended information, where tokens with higher attention scores contribute more to the final output.
The weighted sum operation is akin to focusing on specific parts of the input sequence based on their relevance to the current context. All in all, by incorporating self-attention mechanisms, our Transformer model can capture long-range dependencies in the input RF sequences more effectively.
The Transformer block is parameterized by the following: 1) {\it embed-dim} that is the dimension of the embedding vectors, 2) {\it num-heads} that represents the number of attention heads, 3) {\it ff-dim} which specifies the size of the hidden layer in the feedforward network inside the Transformer block. 
%Transformer layer outputs one vector for each time step of our input sequence. Here, we take the mean across all time steps.
\item {\it Global Average Pooling layer:}
Next, we perform global average pooling across the time dimension (sequence length) of the input sequence, in order to reduce the spatial dimensions of the input sequence to a single vector, aggregating information from all tokens.
%	This step is commonly used to generate a fixed-length representation of variable-length input sequences.
\item {\it Dropout \& Dense Layers:}
Next, we apply dropout regularization with a dropout rate of 0.1 to prevent overfitting by randomly setting a fraction of input units to zero during training. The dropout layers are interlaced between two dense layers with 20 units and 6 units, respectively.  
%Dropout is applied after the global average pooling and dense layers to regularize the model.
\end{itemize}

TABLE \ref{table:trans_arch} provides a compact summary of the architecture of the Transformer model that we have implemented.

{\it Hyperparameters:} 
We set {\it embed-dim} = 32, {\it num-heads} = 4, and {\it ff-dim} = 32.

\begin{table}[t!]
\centering
\begin{tabular}{|c| c| c|} 
 \hline
       {\bf Layer type} &    {\bf Output shape} &   {\bf Parameters} \\\hline
           Input layer &       (None, 16) &     0     \\\hline
           Token \& position embedding &       (None, 16, 32) &     640,416      \\\hline
           Transformer block &       (None, 16, 32) &     19,040      \\\hline
           Global avg. pooling &       (None, 32) &     0      \\\hline
           Dropout &       (None, 32) &     0      \\\hline
           Dense &       (None, 20) &     660     \\\hline
           Dropout &       (None, 20) &     0     \\\hline
           Dense &       (None, 6) &     126     \\\hline
\end{tabular}
\caption{Architecture of our Transformer model. }
\label{table:trans_arch}
\vspace{-1mm}
\end{table}

\subsubsection{The LSTM Model}

We tried various configurations of the Vanilla LSTM model. It turns out that the LSTM that gives the best performance is a relatively shallow LSTM with an input layer, two LSTM layers (with 64,16 units, respectively), each followed by a dropout layer with a dropout ratio of 0.2, followed by one dense layer of size 6. We use Softmax activation function at the final layer. 

{\it Hyperparameters:}    
We use categorical focal crossentropy (with $\gamma=1$) as loss function, we use Adam optimizer. We do 100 epochs, set the batch size to 8192, and use a train-validation-test ratio of 75-15-15.      
%validation_split = 0.10, 

\subsubsection{The CNN Model}

We tried various configurations of the Vanilla CNN model. It turns out that the CNN that gives the best performance is a relatively shallow CNN with three convolutional (CNV) layers (with 64,32,16 filters respectively), one flatten layer, and three fully connected (FC) layers (of size 64,32,6 respectively). Further, the following holds for each CNV layer: kernel size=3, padding=same, stride=1. Note that each CNV layer is followed by a max pooling layer with pool size=2, while each FC layer is followed by a dropout layer with a dropout ratio of 0.3. We use ReLU activation function at each layer, expect the last one where Softmax is used. 

{\it Hyperparameters:}    
We use categorical focal crossentropy (with $\gamma=1$) as the loss function. We use Adam optimizer, we do 100 epochs, set the batch size to 8192, and use a train-validation-test ratio of 75-15-15.

%Remember that the dataset is randomly split into smaller portions for training and testing in the cross-validation approach. For instance, if K is 5, 80 percent of the data would be used for training and the remaining 20 percent will be used for testing and validation. 
% The classifier model will train five times if K=5. 

%%%%%%%%%%%%%%%%%%%%%%%%%%%%%%%%%%%%%%%%%%%%%%%%%%%%%%%%%%%%

%%%%%%%%%%%%%%%%%%%%%%%%%%%%%%%%%%%%%%%%%%%%%%%%%%%%%%%%%%

\section{Results}

We have implemented the aforementioned ML, MLP and DL models on a workstation with the following specs: Intel core i7 3.6 GHz processor with 32 GB RAM. 

\subsection{Performance metrics}

We utilize accuracy, precision, recall, F1-score, support as primary performance metrics to assess the performance of our ML and DL classifiers. Note that $\mathrm{Accuracy}=\frac{\mathrm{Correct \; predictions}}{\mathrm{Total \; observations}}\times 100=\frac{TN+TP}{TN+TP+FN+FP} \times 100$.
Here, $TN$ represents a true negative, $TP$ represents a true positive, $FN$ represents a false negative, and $FP$ represents a false positive. Further, precision $P=\frac{TP}{TP+FP}$, recall $R=\frac{TP}{TP+FN}$, and F1-score=$\frac{2\times (P\times R)}{P+R}$. 
In addition, we also do a performance comparison of the various ML and DL classifiers by means of confusion matrices.
Furthermore, we also do latency analysis of our ML and MLP classifiers in terms of training time (seconds), and prediction speed (observations/sec). 
Finally, we also provide the progression of accuracy and loss functions against the number of epochs, for our DL models.

% and mismatch for breathing
% experiment data are presented in Table 6. The performance analysis of algorithms for
% breathing experiments data are presented in Table 7.
% This section discusses the performance of the proposed SDR-based, ML-empowered non-contact method that captures the subtle but periodic movements of the chest as well as hand for the detection of dehydrated subjects. 

% Additionally, we also construct the confusion matrices in order to investigate the test accuracy of the three ML classifiers. 

\begin{table}
\centering
\begin{tabular}{|l|l|l|l|}
\hline
{\bf ML Algorithms}                                                                     & \begin{tabular}[c]{@{}l@{}} {\bf Accuracy} \\ (\%)\end{tabular} & \begin{tabular}[c]{@{}l@{}} {\bf Prediction} \\ {\bf Speed} (obs/sec)\end{tabular} & \begin{tabular}[c]{@{}l@{}} {\bf Training} \\ {\bf Time} (sec)\end{tabular} \\ \hline
LD                                                            & 75                                                       & 83                                                                  & 4820                                                         \\ \hline
KNN                                                                            & 76.7                                                     & 9.6                                                                 & 9575                                                         \\ \hline
SVM (Linear)                                                                   & 80.7                                                     & 490                                                                 & 7083.9                                                       \\ \hline
SVM (Quadratic)                                                                & 89.4                                                     & 9.3                                                                 & 14984                                                        \\ \hline
SVM (Cubic)                                                                    & 89.6                                                     & 9.1                                                                 & 15728                                                        \\ \hline
\begin{tabular}[c]{@{}l@{}}Ensemble \\ (SD)\end{tabular} & 84.5                                                     & 1500                                                                & 8195.3                                                       \\ \hline
\begin{tabular}[c]{@{}l@{}}Ensemble\\ (BT)\end{tabular}               & 92.6                                                     & 1400                                                                & 590.39                                                       \\ \hline
\begin{tabular}[c]{@{}l@{}}MLP \\ (Narrow)\end{tabular}             & 96.3                                                         & 1200                                                                     &  4485.5                                                            \\ \hline
\begin{tabular}[c]{@{}l@{}}MLP\\ (Medium)\end{tabular}              & 96.7                                                         & 1100                                                                    & 5412.7                                                             \\ \hline
\begin{tabular}[c]{@{}l@{}}MLP\\ (Wide)\end{tabular}                &96.4                                                          & 1000                                                                    & 6560.7                                                             \\ \hline
\begin{tabular}[c]{@{}l@{}}MLP\\ (Bilayered)\end{tabular}           & 95.8                                                         &1300                                                                     & 4265.4                                                             \\ \hline
\begin{tabular}[c]{@{}l@{}}MLP\\ (Trilayered)\end{tabular}          &  95.6                                                        & 1100                                                                    &  5280.2                                                           \\ \hline
\end{tabular}
\caption{Accuracy, prediction speed and training time of various ML and MLP models that we have implemented. }
\label{table:ML_accu}
\vspace{-1mm}
\end{table}

\begin{table}
\centering
\begin{tabular}{|c|c|c|c|c|c|c|c|}
\hline
Algorithms                                                                     & \begin{tabular}[c]{@{}c@{}}Actual /\\ Guess\end{tabular}                   & \multicolumn{1}{r|}{AST}                                                & COPD                                                                       & ILD                                                                        & PN                                                                          & TB                                                                         & N                                                                          \\ \hline
LD                                                            & \begin{tabular}[c]{@{}c@{}}AST\\ COPD\\ ILD\\ PN\\ TB\\ N\end{tabular} & \begin{tabular}[c]{@{}c@{}}\cellcolor{blue!25}75.8\\ 4.7\\ 4.9\\ 4.4\\ 3.6\\ 0.2\end{tabular} & \begin{tabular}[c]{@{}c@{}}2.2\\ \cellcolor{blue!25}68.6\\ 4.4\\ 3.8\\ 2.6\\ 0.1\end{tabular} & \begin{tabular}[c]{@{}c@{}}2.4\\ 4.1\\ \cellcolor{blue!25}66.3\\ 4.3\\ 2.8\\ 0.0\end{tabular}   & \begin{tabular}[c]{@{}c@{}}2\\ 4.7\\ 4.9\\ \cellcolor{blue!25}66.2\\ 3.1\\ 0.1\end{tabular}    & \begin{tabular}[c]{@{}c@{}}2.7\\ 3.8\\ 4.4\\ 4.5\\ \cellcolor{blue!25}73.9\\ 0.3\end{tabular} & \begin{tabular}[c]{@{}c@{}}14.9\\ 14.2\\ 15.1\\ 16.6\\ 13.9\\ \cellcolor{blue!25}99.2\end{tabular} \\ \hline
KNN                                                                            & \begin{tabular}[c]{@{}c@{}}AST\\ COPD\\ ILD\\ PN\\ TB\\ N\end{tabular} & \begin{tabular}[c]{@{}c@{}}\cellcolor{blue!25}80.0\\ 3.4\\ 2.6\\ 2.2\\ 3.4\\ 0.1\end{tabular} & \begin{tabular}[c]{@{}c@{}}1.4\\ \cellcolor{blue!25}71.7\\ 3.1\\ 2.8\\ 1.8\\ 0.1\end{tabular} & \begin{tabular}[c]{@{}c@{}}0.7\\ 1.7\\ \cellcolor{blue!25}66.4\\ 2.0\\ 1.1\\ 0.4\end{tabular} & \begin{tabular}[c]{@{}c@{}}0.5\\ 1.2\\ 2.0\\ \cellcolor{blue!25}65.5\\ 0.5\\ 0.1\end{tabular}  & \begin{tabular}[c]{@{}c@{}}4.4\\ 5.1\\ 6.1\\ 4.7\\ \cellcolor{blue!25}77.1\\ 0.0\end{tabular}   & \begin{tabular}[c]{@{}c@{}}13.0\\ 17.0\\ 19.9\\ 22.8\\ 16.0\\ \cellcolor{blue!25}99.2\end{tabular} \\ \hline
\begin{tabular}[c]{@{}c@{}}SVM\\ (Linear)\end{tabular}                                                                    & \begin{tabular}[c]{@{}c@{}}AST\\ COPD\\ ILD\\ PN\\ TB\\ N\end{tabular} & \begin{tabular}[c]{@{}c@{}}\cellcolor{blue!25}82.6\\ 4.5\\ 1.9\\ 2.4\\ 1.2\\ 0.0\end{tabular}   & \begin{tabular}[c]{@{}c@{}}0.9\\ \cellcolor{blue!25}73.7\\ 2.0\\ 1.6\\ 0.9\\ 0.0\end{tabular}   & \begin{tabular}[c]{@{}c@{}}0.7\\ 2.0\\ \cellcolor{blue!25}74.0\\ 1.2\\ 0.6\\ 0.0\end{tabular}     & \begin{tabular}[c]{@{}c@{}}0.9\\ 1.6\\ 2.6\\ \cellcolor{blue!25}69.4\\ 1.2\\ 0.1\end{tabular} & \begin{tabular}[c]{@{}c@{}}1.8\\ 3.9\\ 1.6\\ 2.2\\ \cellcolor{blue!25}84.5\\ 0.0\end{tabular}   & \begin{tabular}[c]{@{}c@{}}13.1\\ 14.3\\ 18.0\\ 23.1\\ 11.6\\ \cellcolor{blue!25}99.9\end{tabular} \\ \hline
\begin{tabular}[c]{@{}c@{}}SVM\\ (Quad.)\end{tabular}                                                                 & \begin{tabular}[c]{@{}c@{}}AST\\ COPD\\ ILD\\ PN\\ TB\\ N\end{tabular} & \begin{tabular}[c]{@{}c@{}}\cellcolor{blue!25}91.1\\ 3.2\\ 2.7\\ 2.6\\ 1.9\\ 0.1\end{tabular} & \begin{tabular}[c]{@{}c@{}}1.4\\ \cellcolor{blue!25}86.6\\ 1.4\\ 1.5\\ 1.0\\ 0.0\end{tabular}     & \begin{tabular}[c]{@{}c@{}}1.0\\ 1.1\\ \cellcolor{blue!25}84.9\\ 1.3\\ 0.3\\ 0.0\end{tabular}     & \begin{tabular}[c]{@{}c@{}}0.8\\ 0.7\\ 1.6\\ \cellcolor{blue!25}82.6\\ 0.6\\ 0.0\end{tabular}    & \begin{tabular}[c]{@{}c@{}}1.6\\ 2.6\\ 1.9\\ 1.8\\ \cellcolor{blue!25}91.3\\ 0.0\end{tabular}   & \begin{tabular}[c]{@{}c@{}}4.1\\ 5.9\\ 7.6\\ 10.2\\ 4.9\\ \cellcolor{blue!25}99.9\end{tabular}     \\ \hline
\begin{tabular}[c]{@{}c@{}}SVM\\ (Cubic)\end{tabular}                                                                     & \begin{tabular}[c]{@{}c@{}}AST\\ COPD\\ ILD\\ PN\\ TB\\ N\end{tabular} & \begin{tabular}[c]{@{}c@{}}\cellcolor{blue!25}89.3\\ 2.6\\ 1.8\\ 2.0\\ 1.7\\ 0.0\end{tabular}   & \begin{tabular}[c]{@{}c@{}}2.3\\ \cellcolor{blue!25}87.7\\ 1.9\\ 1.7\\ 1.4\\ 0.0\end{tabular}   & \begin{tabular}[c]{@{}c@{}}1.0\\ 1.0\\ \cellcolor{blue!25}86.0\\ 1.5\\ 0.5\\ 0.0\end{tabular}   & \begin{tabular}[c]{@{}c@{}}1.2\\ 0.9\\ 2.1\\ \cellcolor{blue!25}83.8\\ 0.9\\ 0.7\end{tabular}  & \begin{tabular}[c]{@{}c@{}}1.9\\ 2.2\\ 1.1\\ 2.0\\ \cellcolor{blue!25}91.2\\ 0.0\end{tabular}   & \begin{tabular}[c]{@{}c@{}}4.3\\ 5.7\\ 7.0\\ 9.0\\ 4.2\\ \cellcolor{blue!25}99.2\end{tabular}      \\ \hline
\begin{tabular}[c]{@{}c@{}}Ensemble \\ (SD)\end{tabular} & \begin{tabular}[c]{@{}c@{}}AST\\ COPD\\ ILD\\ PN\\ TB\\ N\end{tabular} & \begin{tabular}[c]{@{}c@{}}\cellcolor{blue!25}83.3\\ 0.5\\ 0.7\\ 0.6\\ 0.3\\ 0.0\end{tabular}   & \begin{tabular}[c]{@{}c@{}}0.1\\ \cellcolor{blue!25}82.2\\ 0.0\\ 0.4\\ 0.2\\ 0.0\end{tabular}   & \begin{tabular}[c]{@{}c@{}}0.1\\ 0.2\\ \cellcolor{blue!25}79.9\\ 0.4\\ 0.1\\ 0.0\end{tabular}   & \begin{tabular}[c]{@{}c@{}}0.2\\ 0.5\\ 0.3\\ \cellcolor{blue!25}77.2\\ 0.3\\ 0.1\end{tabular}  & \begin{tabular}[c]{@{}c@{}}0.1\\ 0.4\\ 0.2\\ 0.2\\ \cellcolor{blue!25}84.7\\ 0.0\end{tabular}   & \begin{tabular}[c]{@{}c@{}}16.2\\ 16.1\\ 18.6\\ 21.2\\ 14.5\\ \cellcolor{blue!25}99.9\end{tabular} \\ \hline
\begin{tabular}[c]{@{}c@{}}Ensemble\\ (BT)\end{tabular}               & \begin{tabular}[c]{@{}c@{}}AST\\ COPD\\ ILD\\ PN\\ TB\\ N\end{tabular} & \begin{tabular}[c]{@{}c@{}}\cellcolor{blue!25}93.0\\ 1.7\\ 1.0\\ 1.5\\ 1.0\\ 0.1\end{tabular} & \begin{tabular}[c]{@{}c@{}}0.5\\ \cellcolor{blue!25}91.6\\ 1.1\\ 1.4\\ 0.6\\ 0.0\end{tabular} & \begin{tabular}[c]{@{}c@{}}0.5\\ 0.5\\ \cellcolor{blue!25}88.4\\ 1.4\\ 0.5\\ 0.3\end{tabular} & \begin{tabular}[c]{@{}c@{}}1.6\\ 3.1\\ 4.7\\ \cellcolor{blue!25}89.5\\ 1.5\\ 1.5\end{tabular}  & \begin{tabular}[c]{@{}c@{}}1.7\\ 1.5\\ 1.5\\ 1.3\\ \cellcolor{blue!25}95.0\\ 0.1\end{tabular} & \begin{tabular}[c]{@{}c@{}}2.7\\ 1.6\\ 3.4\\ 4.9\\ 1.5\\ \cellcolor{blue!25}97.9\end{tabular}      \\ \hline
\begin{tabular}[c]{@{}c@{}}MLP \\ (Narrow)\end{tabular}             & \begin{tabular}[c]{@{}c@{}}AST\\ COPD\\ ILD\\ PN\\ TB\\ N\end{tabular} & \begin{tabular}[c]{@{}c@{}}\cellcolor{blue!25}97.1\\ 1.6\\ 1.3\\ 1.6\\ 0.8\\ 0.4\end{tabular} & \begin{tabular}[c]{@{}c@{}}0.7\\ \cellcolor{blue!25}95.7\\ 1.1\\ 1.1\\ 0.6\\ 0.1\end{tabular} & \begin{tabular}[c]{@{}c@{}}0.5\\ 0.6\\ \cellcolor{blue!25}94.6\\ 1.6\\ 0.3\\ 0.1\end{tabular} & \begin{tabular}[c]{@{}c@{}}0.7\\ 0.8\\ 1.4\\ \cellcolor{blue!25}\cellcolor{blue!25}93.4\\ 0.5\\ 0.1\end{tabular}  & \begin{tabular}[c]{@{}c@{}}0.7\\ 0.9\\ 0.9\\ 0.9\\ \cellcolor{blue!25}97.4\\ 0.3\end{tabular} & \begin{tabular}[c]{@{}c@{}}0.4\\ 0.5\\ 0.5\\ 1.4\\ 0.4\\ \cellcolor{blue!25}99.9\end{tabular}      \\ \hline
\begin{tabular}[c]{@{}c@{}}MLP\\ (Medium)\end{tabular}              & \begin{tabular}[c]{@{}c@{}}AST\\ COPD\\ ILD\\ PN\\ TB\\ N\end{tabular} & \begin{tabular}[c]{@{}c@{}}\cellcolor{blue!25}97.3\\ 1.1\\ 1.0\\ 09\\ 0.6\\ 0.3\end{tabular}  & \begin{tabular}[c]{@{}c@{}}0.5\\ \cellcolor{blue!25}95.8\\ 0.8\\ 1.2\\ 0.9\\ 0.0\end{tabular} & \begin{tabular}[c]{@{}c@{}}0.3\\ 0.9\\ \cellcolor{blue!25}95.5\\ 0.9\\ 0.4\\ 0.1\end{tabular} & \begin{tabular}[c]{@{}c@{}}0.5\\ 0.5\\ 1.0\\ \cellcolor{blue!25}95.1\\ 0.3\\ 0.1\end{tabular}  & \begin{tabular}[c]{@{}c@{}}1.0\\ 1.2\\ 1.1\\ 0.8\\ \cellcolor{blue!25}97.4\\ 0.3\end{tabular} & \begin{tabular}[c]{@{}c@{}}0.3\\ 0.5\\ 0.8\\ 1.1\\ 0.4\\ \cellcolor{blue!25}99.3\end{tabular}      \\ \hline
\begin{tabular}[c]{@{}c@{}}MLP\\ (Wide)\end{tabular}                & \begin{tabular}[c]{@{}c@{}}AST\\ COPD\\ ILD\\ PN\\ TB\\ N\end{tabular} & \begin{tabular}[c]{@{}c@{}}\cellcolor{blue!25}96.2\\ 1.2\\ 1.3\\ 1.2\\ 0.9\\ 0.3\end{tabular} & \begin{tabular}[c]{@{}c@{}}1.0\\ 9\cellcolor{blue!25}5.3\\ 1.1\\ 0.9\\ 0.7\\ 0.1\end{tabular} & \begin{tabular}[c]{@{}c@{}}0.7\\ 1.0\\ \cellcolor{blue!25}95.5\\ 0.7\\ 0.6\\ 0.4\end{tabular} & \begin{tabular}[c]{@{}c@{}}0.5\\ 0.7\\ 0.5\\ \cellcolor{blue!25}95.5\\ 0.3\\ 0.1\end{tabular}  & \begin{tabular}[c]{@{}c@{}}0.9\\ 1.3\\ 0.9\\ 0.7\\ \cellcolor{blue!25}97.1\\ 0.3\end{tabular} & \begin{tabular}[c]{@{}c@{}}0.7\\ 0.6\\ 0.7\\ 1.1\\ 0.4\\ \cellcolor{blue!25}98.9\end{tabular}      \\ \hline
\begin{tabular}[c]{@{}c@{}}MLP\\ (Bilayered)\end{tabular}           & \begin{tabular}[c]{@{}c@{}}AST\\ COPD\\ ILD\\ PN\\ TB\\ N\end{tabular} & \begin{tabular}[c]{@{}c@{}}\cellcolor{blue!25}96.6\\ 1.9\\ 1.2\\ 1.4\\ 0.8\\ 0.6\end{tabular} & \begin{tabular}[c]{@{}c@{}}0.7\\ \cellcolor{blue!25}94.2\\ 1.1\\ 1.0\\ 1.2\\ 0.1\end{tabular} & \begin{tabular}[c]{@{}c@{}}0.7\\ 1.4\\ \cellcolor{blue!25}95.0\\ 1.7\\ 0.5\\ 0.2\end{tabular} & \begin{tabular}[c]{@{}c@{}}0.5\\ 1.0\\ 1.4\\ \cellcolor{blue!25}93.8\\ 0.7\\ 0.2\end{tabular}  & \begin{tabular}[c]{@{}c@{}}0.9\\ 1.4\\ 0.8\\ 1.9\\ \cellcolor{blue!25}96.2\\ 0.3\end{tabular} & \begin{tabular}[c]{@{}c@{}}0.4\\ 0.1\\ 0.4\\ 1.1\\ 0.5\\ \cellcolor{blue!25}98.6\end{tabular}      \\ \hline
\begin{tabular}[c]{@{}c@{}}MLP\\ (Trilayered)\end{tabular}          & \begin{tabular}[c]{@{}c@{}}AST\\ COPD\\ ILD\\ PN\\ TB\\ N\end{tabular} & \begin{tabular}[c]{@{}c@{}}\cellcolor{blue!25}96.4\\ 1.1\\ 1.2\\ 1.2\\ 1.0\\ 0.8\end{tabular} & \begin{tabular}[c]{@{}c@{}}0.6\\ \cellcolor{blue!25}94.4\\ 1.5\\ 1.2\\ 1.0\\ 0.2\end{tabular} & \begin{tabular}[c]{@{}c@{}}0.8\\ 1.6\\ \cellcolor{blue!25}94.8\\ 1.7\\ 0.5\\ 0.1\end{tabular} & \begin{tabular}[c]{@{}c@{}}0.7\\ 1.2\\ 1.3\\ \cellcolor{blue!25}93.8\\ 0.4\\ 0.4\end{tabular}  & \begin{tabular}[c]{@{}c@{}}1.2\\ 1.5\\ 0.7\\ 1.4\\ \cellcolor{blue!25}96.7\\ 0.8\end{tabular} & \begin{tabular}[c]{@{}c@{}}0.4\\ 0.2\\ 0.5\\ 0.6\\ 0.5\\ \cellcolor{blue!25}97.8\end{tabular}      \\ \hline
\end{tabular}
\caption{Confusion matrices of various ML and MLP models. (AST:=Asthma, N:=Normal). }
\label{table:ML_perf}
\end{table}

\subsection{Performance of the ML and MLP models }

TABLE \ref{table:ML_accu} summarizes the test accuracy obtained by a number of ML classifiers as well as the MLP classifier. We note that the MLP model along with its variants reports an overall accuracy that is $>95.6$\% and thus outperforms all the ML models. This is followed by the ensemble model with bagged tree (BT) which reports an accuracy of 92.6\%. The SVM (with quadratic and cubic kernels) and ensemble method with subspace discriminant (SD) register an intermediate performance, while the LD and KNN methods perform the worst.
Additionally, TABLE \ref{table:ML_accu} also outlines the prediction speed and training time of all the ML and MLP classifiers. There also, the MLP and its variants distinguish themselves by achieving the highest prediction speeds among all machine learning classifiers, alongside the ensemble BT method, while their training time is comparable with all other ML classifiers.

TABLE \ref{table:ML_perf} provides a comprehensive performance comparison of the various ML and MLP classifiers that we have implemented, in terms of their confusion matrices. 
We make the following observations. 1) All the ML and MLP models report a very high accuracy, i.e., $>97$\%, for the Normal class. This implies that the respiratory performance of a healthy person is remarkably different from that of the patients suffering with pulmonary diseases. This allows even low-complexity ML models to reliably differentiate between a healthy and a sick person. 2) All ML classifiers are able to diagnose Asthma and TB relatively better, compared to other three pulmonary diseases (i.e., COPD, ILD, and PN). This is partly explainable as ILD and PN are the minority classes in the OFDM-Breathe dataset.

\begin{figure*}
        \begin{subfigure}{0.3\linewidth} % Adjust the width as needed
        \includegraphics[width=\linewidth]{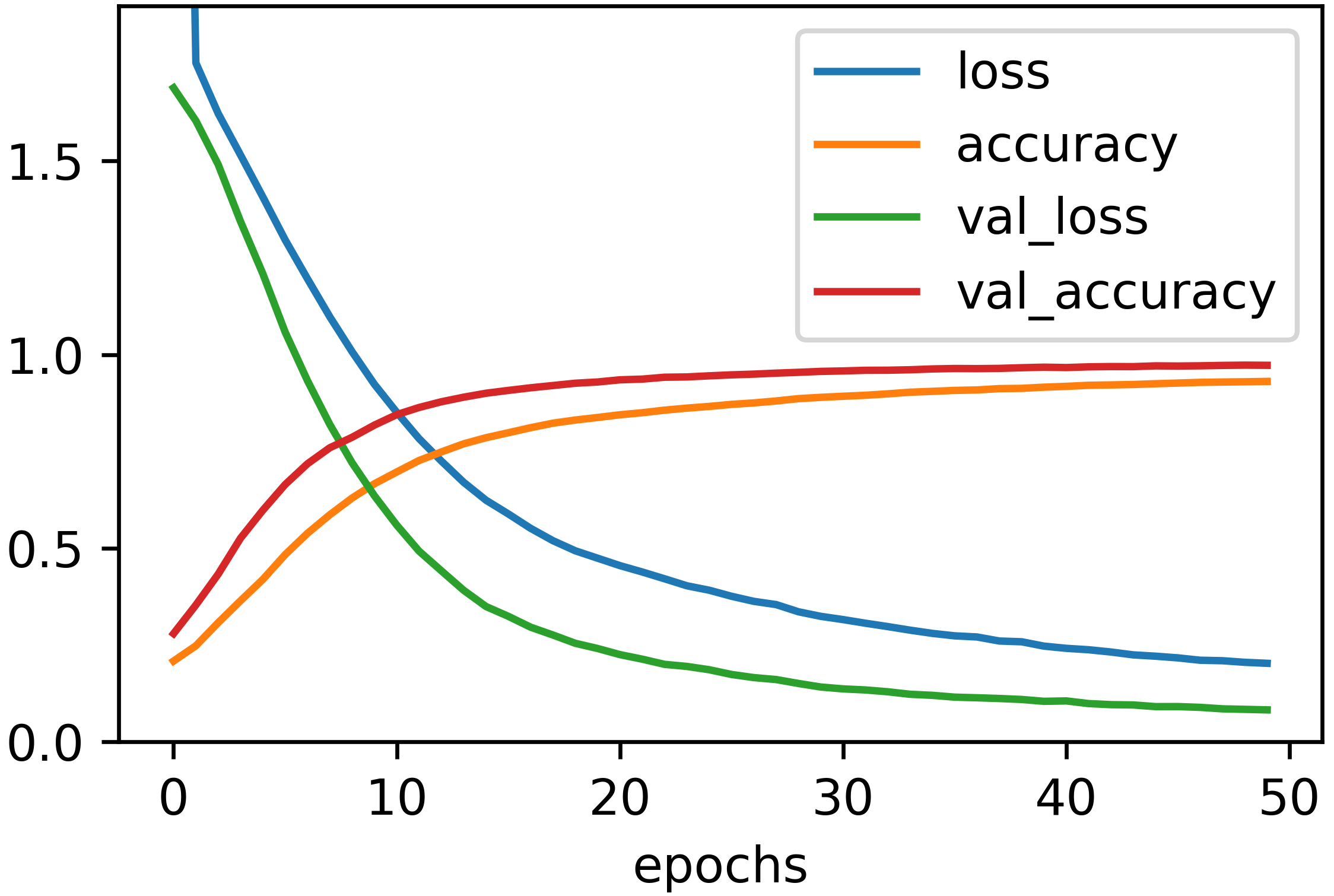}
        \caption{Vanilla CNN.}
        \label{accu_loss_fig1}
    \end{subfigure}
    \hfill
    \begin{subfigure}{0.3\linewidth} % Adjust the width as needed
        \includegraphics[width=\linewidth]{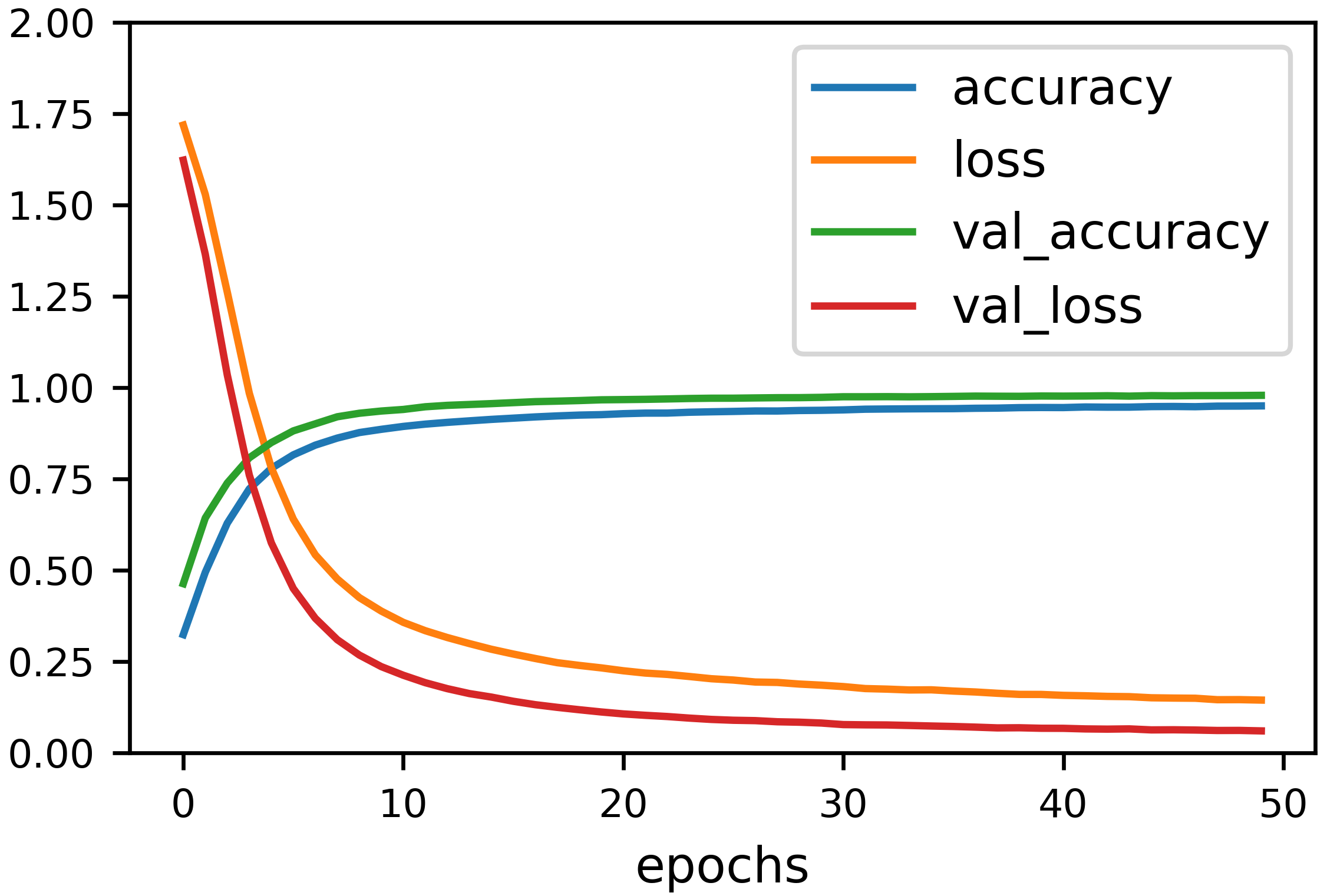}
        \caption{Vanilla LSTM.}
        \label{accu_loss_fig2}
    \end{subfigure}
    \hfill
    \begin{subfigure}{0.3\linewidth} % Adjust the width as needed
        \includegraphics[width=\linewidth,height=1.4in]{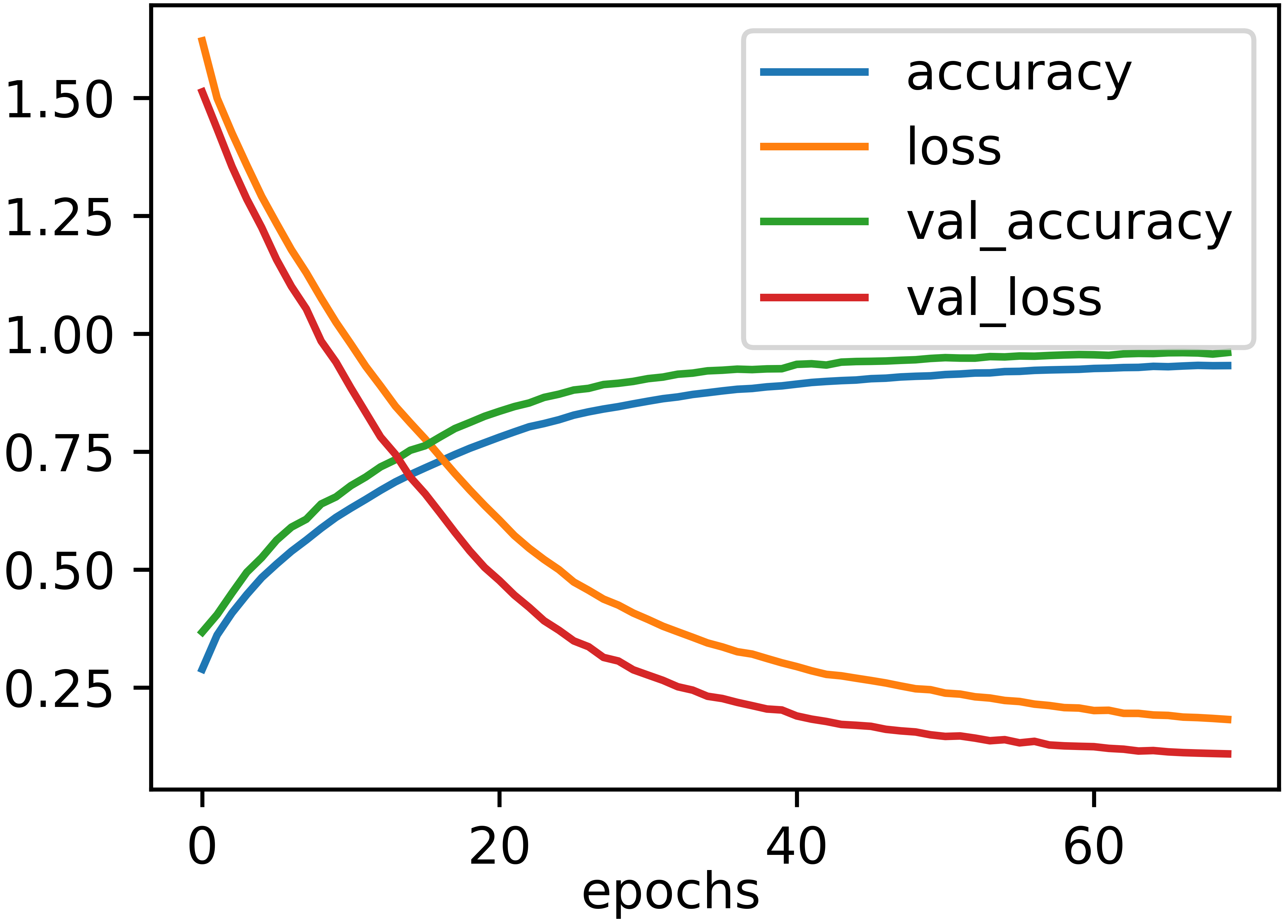}
        \caption{{Vanilla Transformer}.}
        \label{accu_loss_fig3} 
    \end{subfigure}
    \caption{Accuracy and loss functions against the number of epochs, for the three DL models.}
    \label{fig:accu_loss_dl}
\end{figure*}

\begin{figure*}
        \centering
        \begin{subfigure}{0.3\linewidth} % Adjust the width as needed
        \includegraphics[width=2.2in]{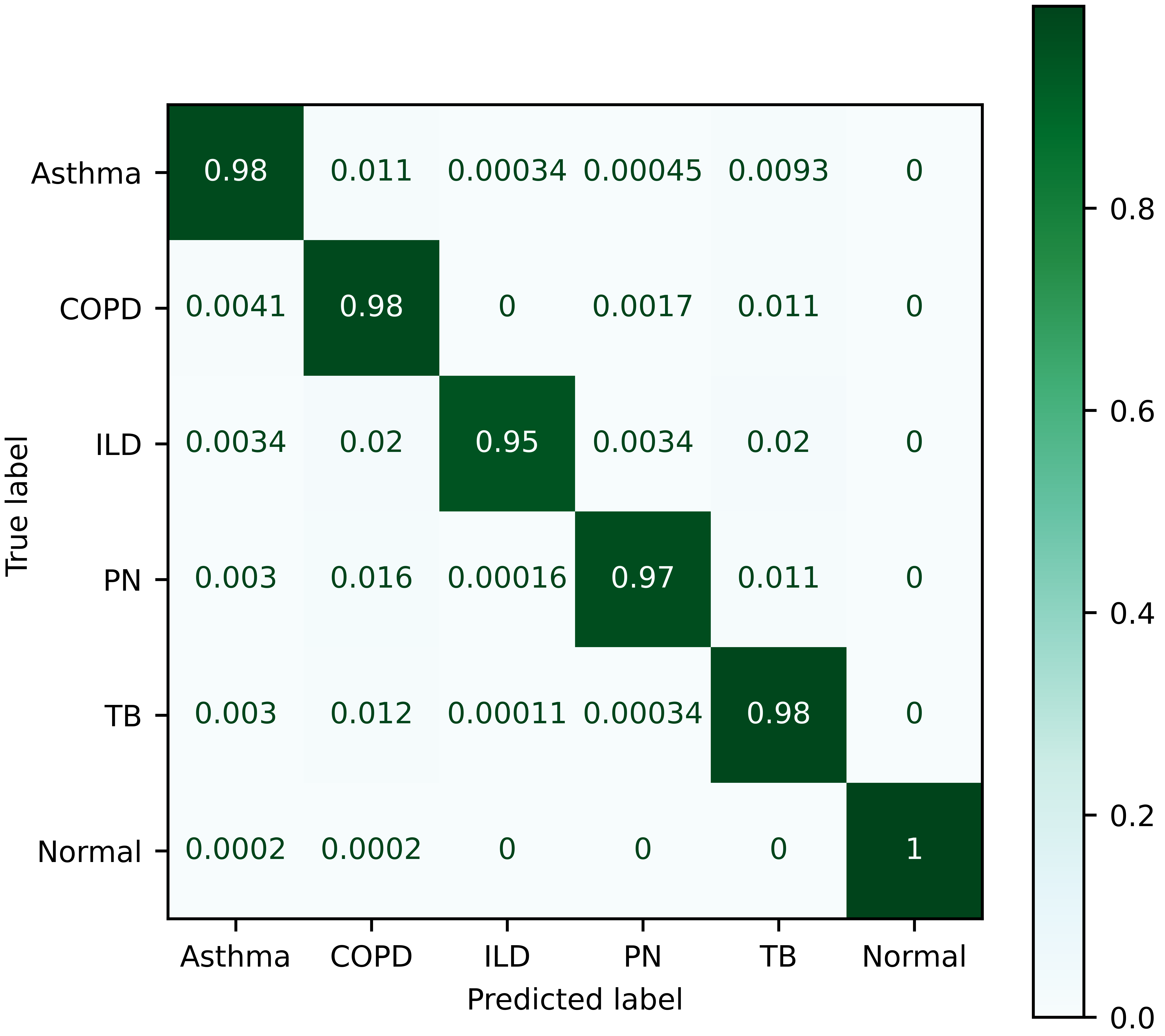}
        \caption{Vanilla CNN.}
        \label{conf_mat_subfig1}
    \end{subfigure}
       \hfill
    \begin{subfigure}{0.3\linewidth} % Adjust the width as needed
        \includegraphics[width=2.2in]{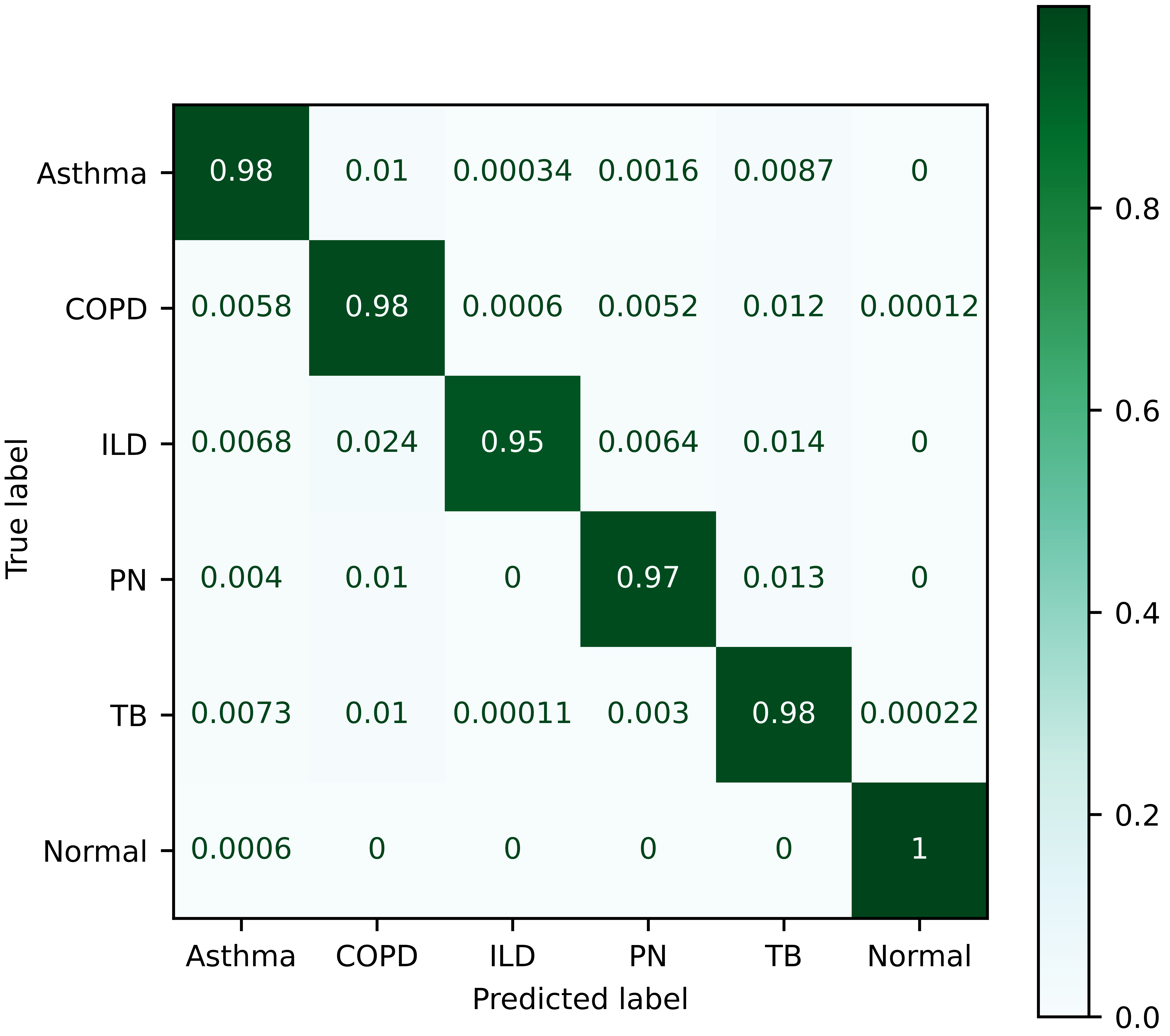}
        \caption{Vanilla LSTM.}
        \label{conf_mat_subfig2}
    \end{subfigure}
    \hfill
    \begin{subfigure}{0.3\linewidth} % Adjust the width as needed
        \includegraphics[width=2.2in]{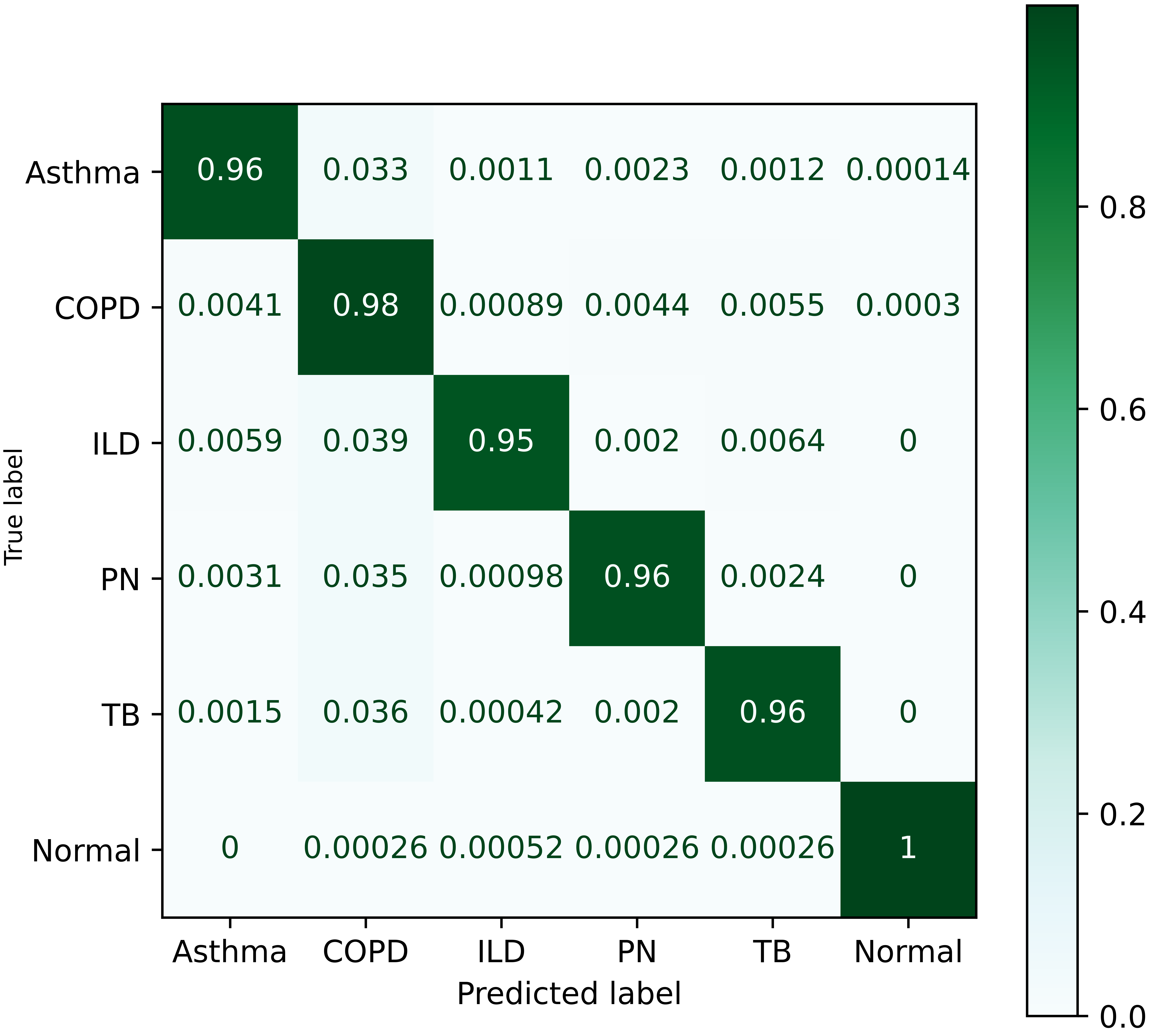}
        \caption{{Vanilla Transformer }.}
        \label{conf_mat_subfig3} 
    \end{subfigure}
    \caption{Confusion matrices for the three DL models.}
    \label{fig:conf_mat_dl}
\end{figure*}

\subsection{Performance of the DL models}

Fig. \ref{fig:accu_loss_dl} plots the accuracy and categorical focal cross-entropy loss function against number of epochs during the training and validation phase, for each of the three DL models, i.e., Transformer, LSTM and CNN. We note that both the loss function and accuracy of the CNN, LSTM and Transformer classifiers saturate to nearly optimal value in roughly 50, 40, 60 epochs, respectively. Fig. \ref{fig:accu_loss_dl} also clearly illustrates that none of the three DL models exhibits any kind of overfitting or underfitting. 

Next, Fig. \ref{fig:conf_mat_dl} plots the confusion matrices of the Transformer, LSTM and CNN models. We observe the following. 1) All three confusion matrices are nearly diagonal matrices, i.e., diagonal-heavy, which implies that all three DL models achieve a very high test accuracy (with a minimal accuracy of 95\% for a given disease class). Specifically, the Transformer, LSTM, CNN models achieve an overall accuracy of 97\%, 97\%, and 98\%, respectively. 2) All the three DL models report an accuracy of 100\% for the Normal class. This is a pleasant result because the respiratory performance of a healthy person is quite distinct from the patients with pulmonary diseases. 3) The CNN model slightly outperforms the other two DL models in diagnosing Asthma with an accuracy of 99\%. 4) All the three DL models report slightly less accuracy of (95-96)\% for the ILD class. This is probably due to the fact that OFDM-Breathe dataset contains data of 5 ILD subjects only.

Table \ref{table:trans_perf} extends further the discussion of Fig. \ref{fig:conf_mat_dl} by providing a comprehensive performance comparison of the three DL models in terms of precision, recall, F1-score and support. Firstly, Table \ref{table:trans_perf} re-affirms the fact that the normal breathing pattern is quite distinct compared to pathological respiratory patterns due to Asthma, COPD, ILD, PN, and TB. Secondly, the Transformer, LSTM, CNN models have the largest precision of 1 for Asthma class, ILD class, ILD \& PN classes, respectively. This implies that the Transformer, LSTM, CNN models never misdiagnose a healthy person (or a person with any other pulmonary disease) with having Asthma, ILD, ILD \& PN, respectively.
Thirdly, the precision (recall) of Transformer and LSTM models for the COPD class is lowest (highest) among all five diseases, i.e., 0.9 (0.99), while the CNN has a high precision of 0.97 and high recall of 0.98 for the COPD class. This implies that both Transformer and LSTM models are biased towards minimization of the false negatives for the COPD class.
Finally, we note that both Transformer and LSTM model have a slightly lower F1-score of 94\%, 95\%, respectively for the COPD class, which makes them slightly less reliable for diagnosing the COPD disease. On the other hand, the CNN registers a very high F1-score of 98\% for all the five disease classes. This implies that the CNN model turns out to be the winner as it strikes a good balance between precision and recall, which in turn makes it suitable for lung disease diagnosis in a clinical setting where both false positives and false negatives have great implications. 
\begin{table}[t!]
\centering
\begin{tabular}{|c| c| c| c| c|} 
 \hline
           \multicolumn{5}{|c|}{\bf The CNN model} \\
           \hline 
           {\bf Class} &  {\bf precision} &   {\bf recall} & {\bf F1-score} &   {\bf support} \\\hline
           Asthma &      0.98 &     0.99 &     0.98 &     8950 \\\hline
           COPD &      0.97 &     0.98 &     0.98 &     8300 \\\hline
           ILD &      1.00 &     0.96 &     0.98 &     2650 \\\hline
           PN &      1.00 &     0.97 &     0.98 &     6300 \\\hline
           TB &      0.98 &     0.98 &     0.98 &     8950 \\\hline
           Normal &      1.00 &     1.00 &     1.00 &     5000 \\\hline
           \hline
           \multicolumn{5}{|c|}{\bf The LSTM model} \\
           \hline 
           {\bf Class} &  {\bf precision} &   {\bf recall} & {\bf F1-score} &   {\bf support} \\\hline
           Asthma &      0.99 &     0.96 &     0.98 &     8950 \\\hline
           COPD &      0.90 &     0.99 &     0.95 &     8300 \\\hline
           ILD &      1.00 &     0.95 &     0.97 &     2650 \\\hline
           PN &      0.99 &     0.96 &     0.98 &     6300 \\\hline
           TB &      0.99 &     0.97 &     0.98 &     8950 \\\hline
           Normal &      0.99 &     1.00 &     1.00 &     5000 \\\hline
           \hline 
           \multicolumn{5}{|c|}{\bf The Transformer model} \\
           \hline 
       {\bf Class} &  {\bf precision} &   {\bf recall} & {\bf F1-score} &   {\bf support} \\\hline
           Asthma &       1.00 &     0.96 &     0.98 &     7249 \\\hline
           COPD &       0.90 &     0.99 &     0.94 &     6759 \\\hline
           ILD &       0.99 &     0.95 &     0.97 &     2045 \\\hline
           PN &       0.99 &     0.96 &     0.98 &     5094 \\\hline
           TB &       0.99 &     0.96 &     0.98 &     7100 \\\hline
           Normal &       1.00 &     1.00 &     1.00 &     3873 \\\hline
\end{tabular}
\caption{Performance of the three DL models. }
\label{table:trans_perf}
\vspace{-1mm}
\end{table}
% \begin{figure*}
%     \centering
%         \begin{subfigure}{0.3\linewidth} % Adjust the width as needed
%         \includegraphics[width=\linewidth]{conf_vanilla_cnn.png}
%         \caption{Vanilla CNN.}
%         \label{conf_mat_subfig1}
%     \end{subfigure}
%        \hfill
%     \begin{subfigure}{0.3\linewidth} % Adjust the width as needed
%         \includegraphics[width=\linewidth]{conf_vanilla_lstm.png}
%         \caption{Vanilla LSTM.}
%         \label{conf_mat_subfig2}
%     \end{subfigure}
%     \hfill
%     \begin{subfigure}{0.3\linewidth} % Adjust the width as needed
%         \includegraphics[width=\linewidth]{conf_vanilla_trans.png}
%         \caption{{Vanilla transformer}.}
%         \label{conf_mat_subfig3} 
%     \end{subfigure}
%     \caption{Confusion matrices for the three DL models.}
%     \label{fig:conf_mat_dl}
% \end{figure*}

\begin{figure}
\begin{center}
    \includegraphics[width=0.5\textwidth]{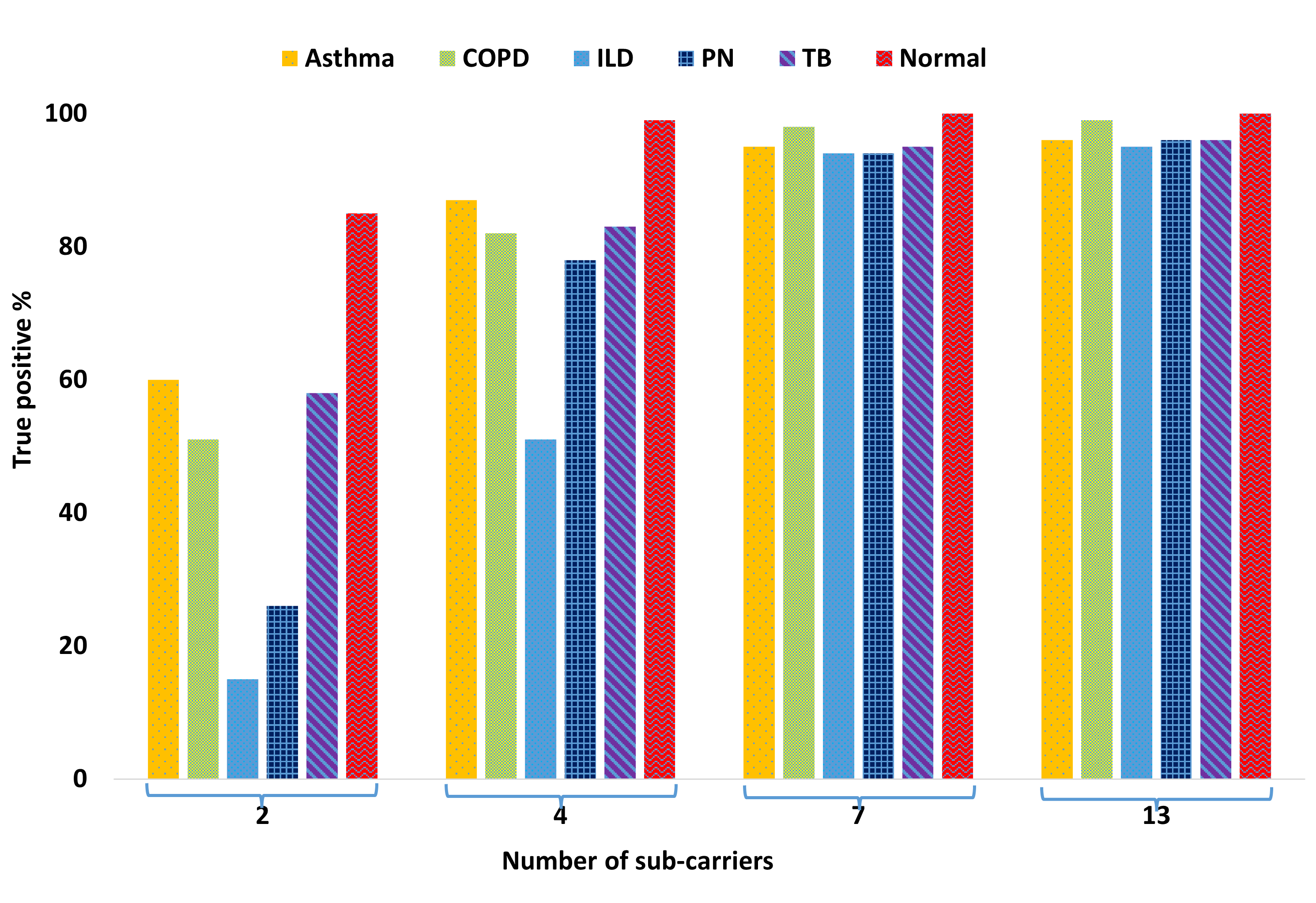}
\caption{{Ablation study to assess the impact of number of sub-carriers on performance of the Transformer classifier.  }}
\label{fig:ablation}
\end{center}
\end{figure}

{\it Ablation study:}
Next, we systematically and monotonically decrease the number of sub-carriers from 13 to 2, in order to quantify how this critical design parameter impacts the performance of our Transformer model. Fig. \ref{fig:ablation} reveals that the Transformer model needs to radio-observe the human chest on at least 7 different microwave frequencies, in order to make a reliable diagnosis (with 96\% accuracy) of the underlying lung disease. Thus, Fig. \ref{fig:ablation} serves as an important benchmark for the 6G ISAC health sensing systems of future that intend to do information exchange and health sensing simultaneously. Precisely speaking, Fig. \ref{fig:ablation} illustrates that the minimum sensing overhead required by our Transformer model to achieve an accuracy of 96\% is 7/64=10.93\% of the allocated bandwidth, which is consumed by ISAC on need basis only. This is quite remarkable because high performance of opportunistic ISAC with very few sub-carriers implies that we have more bandwidth (and thus more data rate) for information exchange.

\section{Conclusion}
\label{sec:conclusion}

We demonstrated a novel non-contact method whereby we radio-exposed the pulmonary patients with microwave OFDM signals in order to diagnose and identify five respiratory diseases, i.e., Asthma, COPD, ILD, PN, and TB. We acquired OFDM-Breathe dataset, first of its kind, which consists of 13,920 seconds of raw RF data at 64 distinct microwave OFDM frequencies, acquired from a total of 116 subjects in a hospital setting (25 healthy control subjects, and 91 pulmonary patients). Among the 91 patients, 25 have Asthma, 25 have COPD, 25 have TB, 5 have ILD, and 11 have PN. We implemented a number of ML and DL models in order to do lung disease classification using OFDM-Breathe dataset. Among all AI models, the CNN model turned out to be the winner in terms of precision, recall, and F1-score. The ablation study revealed that our proposed microwave-based spectroscopy method needs to radio-observe human chest on seven different microwave frequencies only, in order to make a reliable diagnosis of the underlying lung disease. This corresponds to a small sensing overhead of 10.93\%, which points to the feasibility of 6G ISAC systems of future. 

Note that the proposed method merely complements (and not replaces) the existing pulmonary disease diagnosis methods, and thus aims to aid the clinician reach a diagnosis. Future work will investigate the viability of the proposed method for seamless, non-contact monitoring and identification of COVID-19, lung cancer and other respiratory diseases. 

%Finally, this work is also a contribution to the themes of smart homes, smart cities, and digital twin.

%\input{Appendix1}

%\appendices

%\input{appendix}

%\section*{Acknowledgements}

\footnotesize{
\bibliographystyle{IEEEtran}
\bibliography{references}
}

\vfill\break

\end{document}